\documentclass{aa} 

\usepackage{graphicx}
\usepackage[dvipsnames]{xcolor} 
\usepackage{lscape}
\usepackage{float}
\usepackage{txfonts}
\usepackage{color}
\usepackage{soul}

\begin{document} 

\title{VEGAS: A VST Early-type GAlaxy Survey.}
\subtitle{VI. The diffuse light in HCG~86 from the ultra-deep VEGAS images}

\author{Rossella Ragusa\inst{1,2}
	   \and
	Marilena Spavone \inst{1}
          \and
          Enrichetta Iodice\inst{1}
        \and 
	Sarah Brough \inst{3}
        \and 
	Maria Angela Raj \inst{1}
        \and 
	Maurizio Paolillo \inst{2}
        \and 
	Michele Cantiello \inst{4}
        \and 
	Duncan A. Forbes \inst{5}
        \and 
	Antonio La Marca \inst{2}
        \and 
	Giuseppe D'Ago \inst{6}
        \and 
	Roberto Rampazzo \inst{7}
        \and 
	Pietro Schipani \inst{1}
}
   \institute{INAF-Astronomical Observatory of Capodimonte,
 Salita Moiariello 16, 80131, Naples, Italy\\
              \email{rossella.ragusa@inaf.it}
         \and
                  University of Naples "Federico II",via Cinthia 21,Naples 80126, Italy 
        \and
        School of Physics, University of New South Wales, NSW 2052, Australia 
        \and
             INAF-Astronomical Abruzzo Observatory, Via Maggini, 64100, Teramo, Italy 
        \and
        Centre for Astrophysics and Supercomputing, Swinburne University of Technology, Hawthorn, Victoria 3122, Australia 
              \and
              Instituto de Astrofísica, Facultad de Física, Pontificia Universidad Católica de Chile, Av. Vicu\~{n}a Mackenna 4860, 7820436 Macul, Santiago, Chile 
            \and
              INAF $-$ Astronomical Observatory of Padova, Via dell’Osservatorio 8 , I-36012, Asiago (VI), Italy  
              }
  \date{...;...}

\abstract{}{}{}{}{} 
 
  \abstract
   {In this paper we present ultra-deep images of the compact group of galaxies HCG~86 as part of the VEGAS survey.}
   {Our main goals are to estimate the amount of intra-group light (IGL), to study the light and color distributions in order to address the main formation process of the IGL component in groups of galaxies.}
   {We derived the azimuthally averaged surface brightness profiles in the $g$,$r$ and $i$ bands with $g-r$ and $r-i$ average colors and color profiles for all group members. By fitting the light distribution, we have extrapolated the contribution of the stellar halos plus the diffuse light from the brightest component of each galaxy. The results are compared with theoretical predictions.}
   {The long integration time and wide area covered  make our data deeper than previous literature studies of the IGL in compact groups of galaxies  and allow us to produce  an extended 
   ($\sim 160$~kpc) map of the IGL, down to a surface brightness level of $\sim 30$~mag/arcsec$^2$ in the $g$ band. The IGL in HCG~86 is mainly in diffuse form and has average colors of $g-r\sim0.8$~mag and $r-i\sim0.4$~mag. The fraction of IGL in HCG~86 is $\sim16$\% of the total luminosity of the group, and this is consistent with estimates available for other compact groups and loose groups of galaxies of similar virial masses.
   A weak trend is present between the amount of IGL and the early-type to late-type galaxy ratio.
   }
   {By comparing the IGL fraction and colors with those predicted by simulations, the amount of IGL in HCG~86 would be the result of the disruption of satellites at an epoch of $z\sim0.4$. At this redshift, observed colors are consistent with the scenario where the main contribution to the mass of the IGL comes from the intermediate/massive galaxies ($10^{10} \leq M_{*} \leq 10^{11}$~M$_\odot$). }

   \keywords{Galaxies: evolution - Galaxies: photometry - Galaxies: group: general - Galaxies: interactions- intergalactic medium - Galaxies: group: individual: HCG~86  }

   \maketitle
%

\section{Introduction}\label{sec:intro}

In the $\Lambda$-Cold Dark Matter scenario, clusters of galaxies are expected to grow 
over time by accreting smaller groups \citep[e.g.][]{deLucia2006}. 
The intra-cluster light (ICL) is the end product of the material stripped from the galaxy 
outskirts and/or the disruption of dwarf galaxies 
during the infall of galaxies in the potential well of the brightest cluster galaxy 
 \citep[e.g.][]{Rudick_2010,Cui_2013,Contini2014,Montes2014,Jim_nez_Teja_2018,Pillepich2018,DeMaio2018,henden2019baryon,Contini2019,deMaio2020}.
The ICL is therefore a diffuse and very faint component ($\mu_g \geq 27$~mag/arcsec$^2$) that 
grows over time during the infall process \citep{Mihos_2015}. 

In this framework, extended and rich structures of clusters of galaxies form by the assembly 
of smaller elements of groups of galaxies, which have typical virial masses in the range $10^{13} - 10^{14} M_{\odot}$ \citep[e.g.][]{Bower2004}. 
Galaxies spend most of their evolutionary life in groups \citep{Miles2004,Robotham} and 
the intra-group light ~(IGL), which builds during the galaxy interactions and merging in 
these environments, is the precursor of the ICL in clusters of galaxies \citep[e.g.][]{Canas2020}. 
The ICL, as well as the IGL, are therefore key parameters to map the mass assembly history in all dense environments.
Given the low-surface brightness levels involved, this is one of the most challenging 
tasks in the era of deep imaging and spectroscopic surveys.
A great improvement has been possible in the last two decades thanks to the effort to study the low-surface 
brightness (LSB) structures in groups and clusters of galaxies, out to the intra-cluster regions \citep[e.g.][]{Slater_2009,Ferrarese2012,vanDokkum2014,Watkins_2014,Duc2015,Mihos_2015,Fliri_2015,Munoz2015,Merritt2016a,Mihos_2016,Watkins_2016, Trujillo_2016,Mihos2017,DeMaio2018,Huang_2017,Huang_2018,Montes2019,Zhang_2019,deMaio2020,montes2021buildup,Delgado21}.
The {\it VST Early-type GAlaxy Survey} 
(VEGAS\footnote{Visit the website \url{http://www.na.astro.it/vegas/VEGAS/Welcome.html}})
has played a pivotal role in this field. 
By combining the large field of view of OmegaCAM@VST and long 
integration time, VEGAS data allow us to map the surface brightness of galaxies down 
to $\mu_g \sim30$~mag/arcsec$^2$
and out to about 10 effective radii ($R_e$), to estimate the ICL/IGL and therefore to relate galaxy structure to the environment
\citep[see][and references therein]{Iodice2017a,Spavone2018,Cattapan2019,Iodice2019,Iodice_2020,Spavone_2020,Raj_2020}.

In this paper we present new deep images of the Hickson Compact Group HCG~86, as part of the VEGAS sample. 
 Hickson compact groups (HCGs) host from four to ten very close galaxies \citep{1982ApJ...255..382H}, with 
low velocity dispersion ($\sim 200$ km/s) and angular size (i.e. angular diameter of the circle containing 
the group) in a range of $0.7-16.4$~arcmin.
The compact configuration makes HCGs among the best sites to study the IGL, 
since the frequent galactic interactions and intense stripping 
are very efficient mechanisms in the build-up of the diffuse light component.
The most recent studies of IGL in HCGs estimated an IGL fraction ranging from 
0\% up to 46\%, 
compared to the total light of the group  and a maximal extension of  80 kpc from the group center \citep{DaRocha2005,DaRocha2008}. 
HCG~79 is the most compact group in the Hickson catalogue \citep{Hickson1992} and 
\citet{DaRocha2005} find a very large amount of IGL in this group, about 46\% 
of the total group light. The IGL component presents an irregular shape and there are
signs of strong past interactions between group members, which induced bars, 
tidal tails and dust lanes. They conclude that this group is in a very advanced stage of evolution, close to collapse into a single structure. 
In contrast no IGL component was detected in HCG~88, formed by four 
late-type galaxies \citep{DaRocha2005}. In the group 90\%  of the neutral HI gas 
is still associated with the galaxy disks \citep{Verdes_Montenegro_2001}, 
suggesting that it is in an early phase of its evolution. 
According to the IGL fraction and galaxy morphologies, the other HCGs analyzed in previous 
works seem to be at intermediate stages of their evolution \citep{DaRocha2005,DaRocha2008}.
HCG~95 presents a spherical IGL component that corresponds to about
10\% of the total light of the group, and to about one third of the total light of HCG~95A, 
the brightest group galaxy (BGG). 
Since there are clear signs of interactions between HCG~95A and HCG~95C, the 
IGL probably formed by the stripping of material from the interacting galaxies \citep{DaRocha2005}.
In HCG~15, HCG~51 and HCG~35 the IGL fraction is very similar, at 19\%, 26\% and 15\% 
of the total light of the group. These components have irregular morphologies, suggesting that these HCGs are far from being relaxed and virialized structures \citep{DaRocha2008}.
Recently, \citet{Poliakov2021} studied the IGL in a large sample of HCGs. Authors found that the 
average surface brightness for IGL is in a range of $25.3 < \mu_r< 28.3$ mag/arcsec$^2$ for all groups in the sample, 
and the fraction in five groups ranges from $7.5\%$ to $25.1\%$. 
They concluded that the mean surface brightness of the IGL depends on the total luminosity of the group 
and becomes brighter in the groups with a larger fraction of early-type galaxies.

The Hickson compact group HCG~86 is a quartet of early-type galaxies (ETGs), with a 
virial mass of $M_{vir}=8.51 \times 10^{12} M_{\odot}$ \citep{Coziol_2004},
located at a distance of 81.73 Mpc \citep{Hickson1992}. 
{The mass-to-light ratio of the group in the K-band is 50, and the average velocity dispersion of the group members is 368 km/s 
\citep[see Tab.~\ref{tab:2MASS}][]{D_az_Gim_nez_2012}. According to
\citet{Proctor_2004}, the two brightest members of the group, HCG~86A  and HCG~86B, are 
$\sim$ 12.8 Gyr and $\sim$ 9 Gyr old, respectively, based on ages and metallicities of unresolved (integrated light) stellar populations, measured from Lick system analysis.
\citet{Ribeiro1998} analyzed the structural and dynamical properties of HCG~86 in a region of $0.5^{\circ} \times 0.5^{\circ}$ around the group. 
They found that HCG~86  has the most members and is the most complex group of 
their catalogue, since it consists of two different kinematic structures, centred on the 
BGG HCG~86A and on HCG~86B}. Based on their analysis, the authors proposed two possible scenarios 
for group formation: it could be a bimodal structure in a merging process, or it is a single 
group, not relaxed yet, where HCG~86B is an extension of HCG~86A in the velocity space. 
In the latter case, considering the numerous dwarf galaxy members  \citep[$\sim 18$,][]{Ribeiro1998}, HCG~86 turns to 
have the most members and be the complex group in the HCGs catalogue. 
On a larger scale, HCG~86 seems to be part of a loose cluster of galaxies, which is 
not dynamically relaxed. The particular configuration of HCG~86 can give key insights into the formation of IGL. 
The new multi-band images presented in this work represent 
the deepest data available to study the IGL in the HCG environment.
This pilot work shows what can be achieved in low surface brightness (LSB) analysis for groups of galaxies.

This work is organized as follows. In Sec.~\ref{sec:obs} we present the observations and the data reduction. In Sec.~\ref{sec:phot} we describe in detail the method used for the data analysis (i.e. the surface photometry). In Sec. ~\ref{sec:igl} we present the results on the intra-group light in HCG~86 and in Sec. ~\ref{sec:concl} we compare our results with the previous ones presented in the  literature, both on the observational and theoretical side. In Sec.~\ref{conc} we summarise the results obtained in this study and draw conclusions.

We have adopted a distance for HCG~86 of 81.73 Mpc. 
To estimate all the distances of this work we used the heliocentric radial velocity (see Tab.~\ref{tab:sample}), given by NED (NASA IPAC Extragalactic Database), and 
$H_0$ = 73 km s\textsuperscript{-1} Mpc\textsuperscript{-1} \citep[][]{Riess2018}. 
Therefore, 1 arcsec corresponds to $\sim$ 0.4 kpc. 
The magnitudes through out the paper are provided in the AB system, and are corrected for Galactic extinction using the extinction coefficients provided by \citet{Schlafly_2011}.

\begin{table*}
\setlength{\tabcolsep}{1.5pt}
\centering
\caption{Properties of the HCG~86 group.} 
\begin{tabular}{lccccccccccccccc}
\hline\hline
     ID &  R.A &    Decl. & v &  N  & $K_b$ & $\mu_K$ &   $\theta_G$    &  $R_{i,j}$ & b/a  & $\sigma_v$ & H$_0$t$_{cr}$ & $M_{VT}$/$L_K$ & $M_{vir}$ & $R_{vir}$\\
     (1) & (2) &(3) &(4) &(5) &(6) &(7) &(8) &(9) &(10) &(11)&(12) &(13) &(14) &(15) \\ 
     & [J2000] & [J2000] &[km/s]&  & & [mag arcsec$^{-2}$] & [arcmin] & [kpc h$^{-1}$]&  &[km/s]&& [h $M_{\odot}$/$ L_{\odot}$] & [$ M_{\odot}$] & [kpc] \\
    
     \\
     HCG~86 & 19:51:59 & - 30:49:31 &5891 &  4 & 9.47 & 20.22 & 3.90 & 46.60 & 0.80& 368 & 0.011& 50& 8.51 x $10^{12}$ & 410\\
    \hline
    \end{tabular}
    \tablefoot{Col.1: Group ID. Col.2 and Col.3: right ascension and declination of the  HCG~86 centre.
    Col.4: median velocity of the group. Col.5: number of galaxy members in the HCG~86 within 3 mag of the brightest member. Col.6 and 7: Galactic extinction-corrected K-band apparent magnitude of the brightest galaxy and the Galactic extinction-corrected K-band group surface brightness. Col.8: angular diameter of the smallest circumscribed circle around group members. Col.9: median projected separation among galaxies. Col.10: apparent group elongation. Col.11: radial velocity dispersion of the galaxies in the HCG~86 computed using individual galaxy velocities. Col.12: dimensionless crossing time. Col.13: mass-to-light ratio in the K band. Col.14 and 15: virial mass and virial radius of the group \citep{Coziol_2004}.}
    \label{tab:2MASS}
\end{table*}

\begin{figure*}[h]
    \centering{}
    \includegraphics[width=18cm]{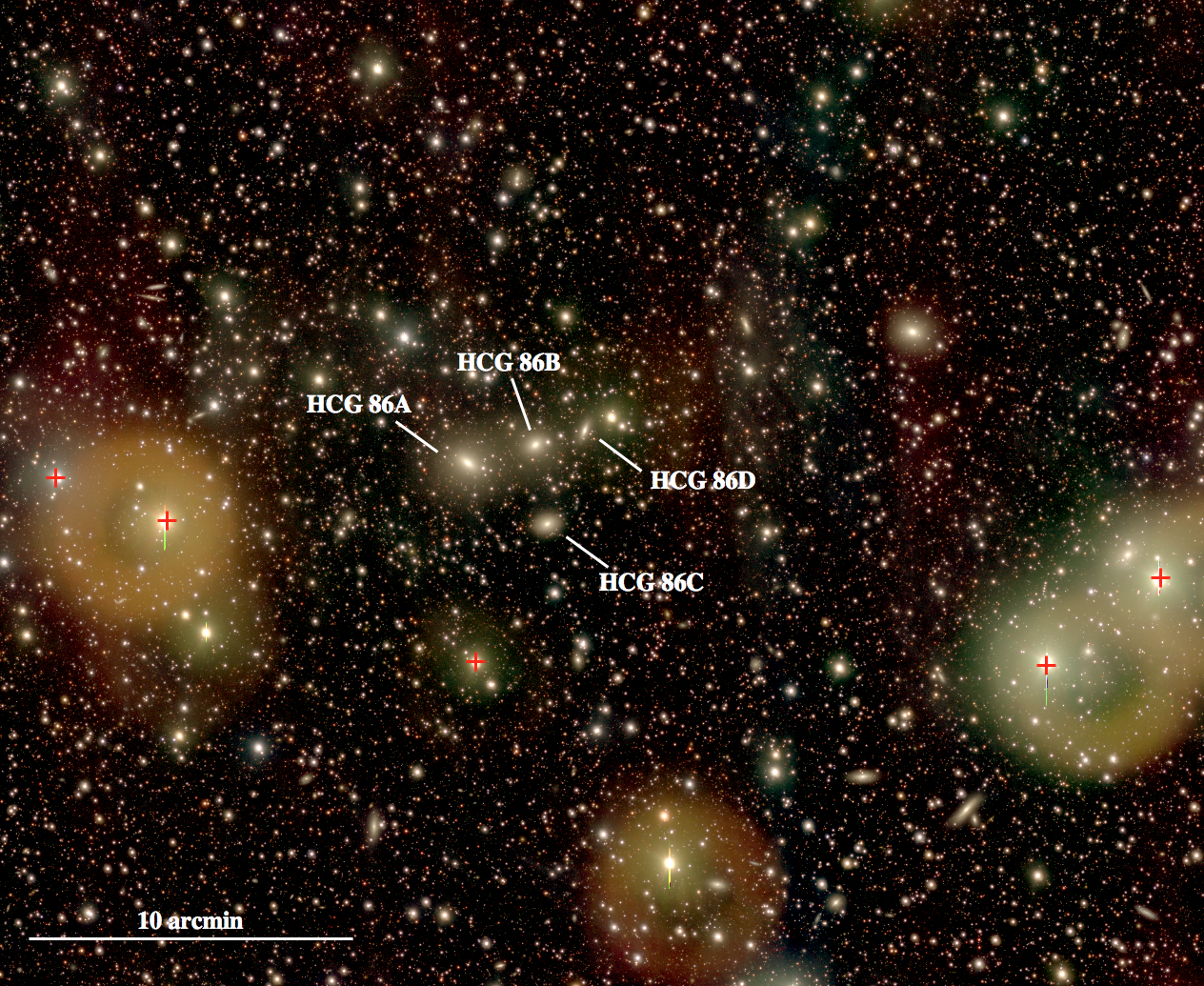}
    \caption {Color composite ($gri$) VST image of the compact group HCG86. The image size is $0.62 \times 0.51$~degrees.  
    North is up and East is to the left. The brightest group members are labelled on the image. The red crosses mark the
    stars that were modelled and subtracted from the image before the analysis of the galaxies.
    See Sec.~\ref{sec:phot} for details.}
    \label{fig:col_comp}
\end{figure*}

\section{Deep images of HGC~86: observations and data reduction}\label{sec:obs}

The data presented in this work are from the VEGAS survey.
VEGAS is a multi-band {\it u}, {\it g}, {\it r} and {\it i} imaging survey carried out with the European Southern Observatory (ESO) Very Large Telescope Survey Telescope (VST). 
The VST is a 2.6~m wide field optical telescope \citep{Schipani2012} equipped with OmegaCAM, a $1^{\circ} \times 1^{\circ}$ camera with a resolution of $0.21$~arcsec~pixel$^{-1}$. 
The data we present in this work were acquired in visitor mode (run ID: IDs:103.A-0181A), 
during dark time in photometric conditions, with an average seeing of FWHM$\sim 0.99$~arcsec in the $g$ band, FWHM$\sim 0.65$~arcsec in the $r$ band and FWHM$\sim 1.4$~arcsec in the $i$ band. 
The $g$-band image is the deepest one, with an exposure time of 5 hours. 
The total integration time in the $r$ and $i$ bands are 3.25 hours and 2.05 hours, respectively. 
Observations were acquired adopting
the standard diagonal dithering strategy. As described in \citet{Capaccioli2015} and \citet{Spavone2017b}, for the targets observed with this strategy, the background subtraction is performed by fitting a surface, typically a 2D polynomial, to the pixel values of the mosaic that are unaffected by celestial sources or defects. 

All the data were processed using the dedicated {\it AstroWISE} pipeline developed to reduce OmegaCam observations \citep{McFarland2013, Venhola2018}. 
The various steps of the AstroWise data reduction were extensively described in \citet{Venhola_2017,Venhola2018}. 

In Fig.~\ref{fig:col_comp} we show the resulting sky-subtracted color composite VST image obtained for HCG~86.
The surface brightness depths at $5 \sigma$ over the average seeing area of 
FWHM=1.26 arcsec, are $\sim$  
$\mu_g=30$~mag/arcsec$^2$, $\mu_r=29$~mag/arcsec$^2$ and $\mu_i=28$~mag/arcsec$^2$ in the $g$, $r$ and $i$ band respectively. 

\section{Data Analysis: surface photometry}\label{sec:phot}
 
\begin{figure*}
\centering

\includegraphics[width=18cm]{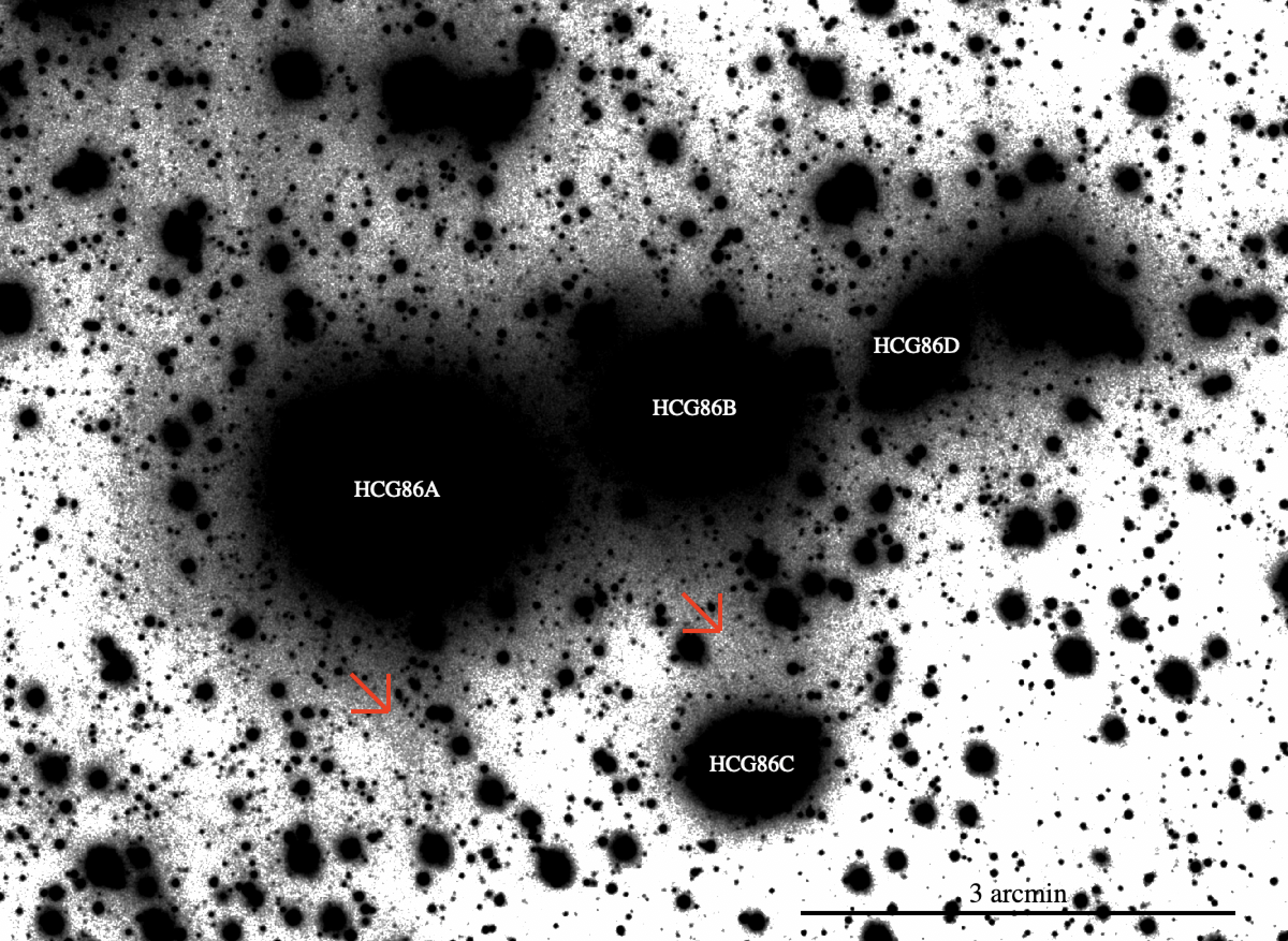}
\caption{ Enlarged region of the deep VST image in the $g$ band centred on the HCG~86 group, of $8.90 \times 6.49$~arcmin. The two red arrows indicate the faint low-surface brightness features contributing to the IGL. The brightest group members (HCG~86A, HCG~86B, HCG~86C and HCG~86D) are also marked in the image.
}
\label{fig:HCG86_dss}
\end{figure*}

\begin{figure*}[h]
   \centering{}
    \includegraphics[width=18cm]{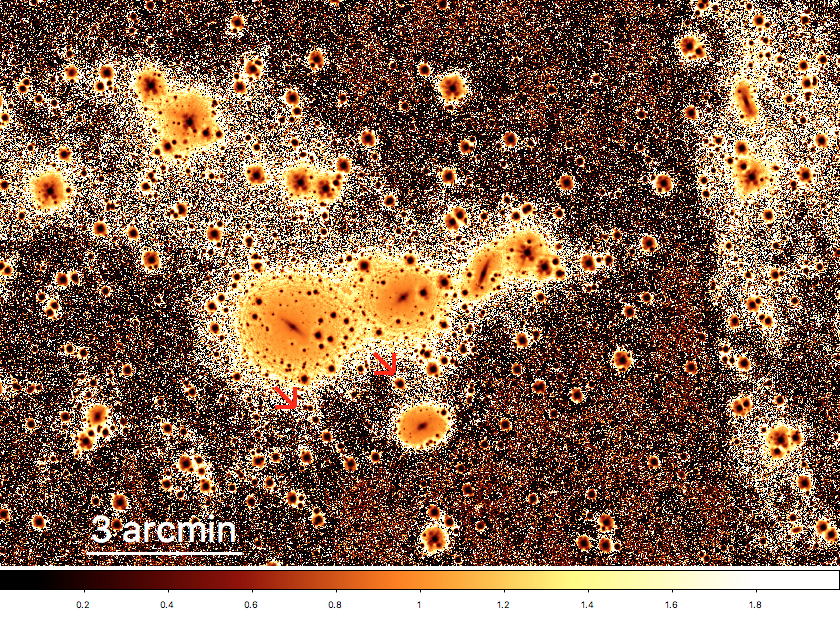}
    \caption{Unsharp-masked image of HCG~86 with gaussian smoothing (radius = 4 pixels). It results from the ratio of the VST $g$-band image and its FMEDIAN smoothed version, using a smoothing box of $50\times50$ pixels. The colorbar indicates the intensity values of the ratio. The image shows an extended envelope of diffuse light associated with the group and symmetrically distributed around the galaxies. The two red arrows indicate the faint low-surface brightness features: the bridge (between HCG 86C and the group) and the filamentary structure at S-E.}
    \label{fig:fmedian}
\end{figure*}

 Fig.~\ref{fig:HCG86_dss} shows an enlarged region of the deep VST image in the $g$ band centred on the 
HCG~86 group.
The deeper VST data shows a large amount of diffuse light, mainly located 
in the galaxies' envelopes. The diffuse light is at least two times more extended than the brightest central regions of the group members. In addition, in the Southern part of the group, we 
detect a faint stellar bridge ($\mu_g \sim 27.8 - 29.1$ mag/arcsec$^2$) connecting HCG~86C with the group center of 
about 1.2 arcmin long and 1 arcmin wide, and two faint streams ($\sim 1.2$ arcmin long)
protruding from the SE regions of HCG~86A (these are marked with red arrows in Fig.~\ref{fig:HCG86_dss}).
There is also an extended region of diffuse light in the North of the group located about 4 arcmin from the group centre, but this 
could be due to light emitted from Galactic cirrus, which contaminates the area (see details in Sec.~\ref{sec:cirrus}).
Similar features stand out from the unsharp-masked $g$-band image, shown in Fig.~\ref{fig:fmedian}. 
This have been obtained using the FMEDIAN task in IRAF, with smoothing boxes of 
$50\times50$ pixels, and taking the ratio of the $g$-band image to its FMEDIAN smoothed version. The image shows the extended envelope of diffuse light associated with the group, which is
symmetrically distributed around the galaxies, and the cirrus emission on the West and North-East sides. 
The bridge connecting the group with HCG~86C and the South-East filamentary structure are also well defined in shape.
The small smoothing box allow us to reveal the internal structure of the group members:
HCG~86A has a disk in the central region, a bar-like feature appears to be present in both HCG~86B and HCG~86C, and
the disk in HCG~86D appears warped in the outskirts.

The main goal of this work is to estimate the diffuse light in the intra-group region.
To this aim, two main steps are needed in the data analysis: 
{\it i)} the contamination from all sources (foreground stars, background galaxies and cirrus emission) 
that contribute to the light must be carefully taken into account; {\it ii)} the light distribution from the
bright group members must be modelled and subtracted from the total budget of the emitted light associated with the group
(i.e. galaxy light plus IGL).
In the following sections we describe the tools and methods adopted in each step. They have all been successfully applied to the previous VST images and optimised to detect and study the LSB features, including the ICL, in several already published papers 
\citep[see][]{Iodice2017a,Spavone2018,Cattapan2019,Iodice2019,Iodice_2020,Spavone_2020,Raj_2020}.

\subsection{Galactic Cirrus emission in region of HCG~86}\label{subsec:cirrus_mask}

\begin{sidewaysfigure*}

    \includegraphics[width=12cm]{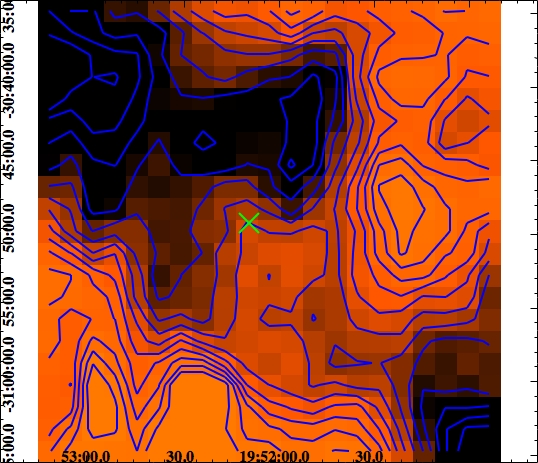}
    \includegraphics[width=12cm]{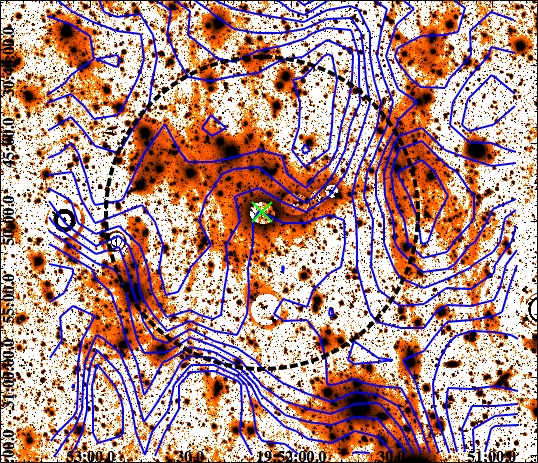}
    \caption{{\it Left panel}- Cirrus map derived from the 100$\mu m$ image available at the 
    NASA/IPAC Infrared Science Archive, with superimposed contours (blue) in the flux interval from $9.25-9.78~ MJy/sr$. 
    {\it Right panel} - Residual image in the $g$ band, where the brightest regions of the group members were subtracted from the original image. The image is $35\times30$~arcmin. The blue lines are the cirrus intensity levels shown in the left panel. The dashed black circle indicates the circular region of $R=600$~arcsec. The green cross indicates the centre of the HCG~86 group in both panels. The residual image shows an extended envelope of diffuse light symmetrically distributed around the group, as already pointed out from unsharp-masked image of HCG~86 (see Fig.~\ref{fig:fmedian}).}
    \label{fig:cirrus_IGLmap}
\end{sidewaysfigure*}

The Galactic cirrus emission is a non-negligible source of contamination in 
LSB imaging. In the optical, the Galactic cirrus emission is due to the starlight belonging to the Milky Way, 
diffused by dust in the local interstellar medium, and then re-emitted in the infrared. 
Therefore, images in the infrared can be used to evaluate the importance of this contamination.
In deep optical imaging, such emission creates pseudo-structures that resemble LSB 
features, like tidal tails or stellar streams \citep{Cortese_2010,mihos2019deep,Duc2015}.
In Fig.~\ref{fig:HCG86_dss} many filamentary structures typical of cirrus are clearly visible.
The most prominent is in the West, extending North-South for about $\sim20$~arcmin.
From the NASA/IPAC Infrared Science Archive we derived the 100$\mu m$ map in the region
of the HCG~86 group, shown in Fig.~\ref{fig:cirrus_IGLmap}.
The 100$\mu m$ map confirms that light from cirrus dominates in the filamentary structure in the West and also 
in the South-East of the group. 
Lower emissions are observed North/North-East and to the South-West of the group. 

The lower spatial resolution of the IRAS map does not allow us to resolve all of the features visible in the VST optical images.

The contribution of the cirrus emission to the diffuse light in HCG~86 will be discussed in detail in  
Sec.~\ref{sec:cirrus}. At this step of the analysis, we derive the intensity contours of the Galactic 
cirrus emission from the 100$\mu m$ map (see left panel of Fig.~\ref{fig:cirrus_IGLmap}) to build an accurate mask\footnote{The mask has been created by using the IRAF task MSKREGIONS.}  of all the regions contaminated by the cirrus emission. This is used in the fitting of the light distribution of stars (Sec.~\ref{subsec:masking}) 
and galaxies (Sec.~\ref{subsec:fitgal}) 
in order to exclude this contribution from the analysis. 

\subsection{Removal of the scattered light from the bright stars}\label{subsec:masking}

In order to remove the contamination of the scattered light from the brightest stars in the field, 
we performed a 2-dimensional (2D) fit of the light distribution by adopting a circularly symmetric 
model for each of them. 
This is based on the isophote fit using the IRAF task ELLIPSE (fixing the centre, position angle and ellipticity),
mapping the light almost out to the edge of the frame, i.e. $\sim$ 33 arcmin.
This fit was carried out after having carefully masked the core of the group, all background and foreground sources, and all the regions affected by the Galactic cirrus emission (see Sec.~\ref{subsec:cirrus_mask}).
In order to preserve the light associated with the group, including the faint emission from the IGL, we
adopted a circular mask centred on the brightest group member HCG~86A with a radius of 13.5 arcmin.
This value corresponds to $\sim$ 0.8 the virial radius of the group (see Tab.~\ref{tab:2MASS}).

We derived a 2D model of each star using the IRAF task BMODEL, and this was then subtracted from the image.
Six stars are modelled in the field\footnote{The modelled stars are the following: 
TYC 7439709-1 R.A.=19:52:07 and DEC=-30:55:35.68 with $m_B=12.57$~mag, 
HD 187783 R.A.=19:52:52 DEC=-30:51:18 with $m_B=10.71$~mag, 
CD 3117130 R.A.=19:53:08 DEC=-30:49:57  with $m_B=10.73$~mag, 
HD 187368 R.A.=19:50:45 DEC=-30:55:43.6 with $m_B=9.42$~mag,
HD 187309 R.A.=19:50:29 DEC=-30:53:03  with $m_B=9.48$~mag,
R.A.=19:51:48.2 and DEC=-30:48:07.8.},  
two on the East side, two on the West side and one more on the South side of HCG~86C (marked in Fig.~\ref{fig:col_comp}), all brighter than $m_B=9.48$~mag. 
The star next to HCG~86D was also modelled and subtracted to prevent its scattered light from affecting the light profile of the galaxy.
Since the 2D model we performed for the brightest stars is symmetric, it cannot account for the bright ghosts that are 
present around them. They are due to the asymmetry of the point-spread functions, which depends on their position in the 
camera. However, they do not contribute to the scattered light at larger distance from each stars, which is the
contribution that we aim at estimating for the purpose of this work.
Therefore, once the symmetric model is subtracted from the parent image, the residuals around each star and
the ghosts are masked.

\subsection{Estimate of the limiting radius and isophote fitting}\label{subsec:fitgal}

To derive the azimuthally averaged surface brightness profiles for the brightest 
group members, in all bands, we used the method from \citet[][]{Pohlen2006}, 
which has also been used for the analysis of VST data by \citet[][]{Iodice2016} 
and in all subsequent papers based on VEGAS data. This is based on two steps: 1) estimate of the limiting radius R$_{lim}$, and 2) fit of the isophotes out to R$_{lim}$ for each galaxy of the group. 

The limiting radius R$_{lim}$ sets the limit of the data, where the galaxy's light blends into the 
background fluctuations and the signal-to-noise ratio (S/N) is about one. 
Since the images are sky-subtracted, the background fluctuations 
are the deviations with respect to the average sky value.
To estimate R$_{lim}$ and the background fluctuations (which also provide an estimate of the accuracy of
the sky-subtraction and flat-fielding) 
we have performed the fit of the isophotes, of the brightest group members HCG~86A and of
HCG~86C (which maps a different region of the image) out to the edge of the star-removed image, in each band.
According to \citet[][]{Pohlen2006}, this is done over elliptical 
annuli (i.e. fixing the ellipticity and position angle), by using the IRAF task ELLIPSE with 
a median sampling and k-sigma clipping algorithm for cleaning deviant sample points at each annulus.
The extensive experience acquired with previous works, shows
that this approach (i.e. combining the median sampling and sigma clipping rejection algorithm) 
improves the isophotal fitting.
Since all the other bright group members (HCG~86B and HCG~86D) are close in space to HCG~86A, 
we can reasonably assume the same value of R$_{lim}$ derived for HCG~86A.
At this step, we made two new masks. One is needed to exclude from the fit all sources outside the group region 
(including previous identified background and foreground sources, plus the cirrus emission and residuals around the 
bright stars, that were subtracted from the image, see Sec.~\ref{subsec:cirrus_mask} and Sec.~\ref{subsec:masking}) 
and all the other group members except HCG~86A. For each galaxy, the masking process is very accurate in order to
take into account the symmetry of the object and the region overlapping with HCG~86A. 
The second mask is adapted to exclude also the group center, and allow us to perform the
fit centered on HCG~86C.

\begin{figure*}
   \resizebox{.5\textwidth}{.4\textheight}{\includegraphics{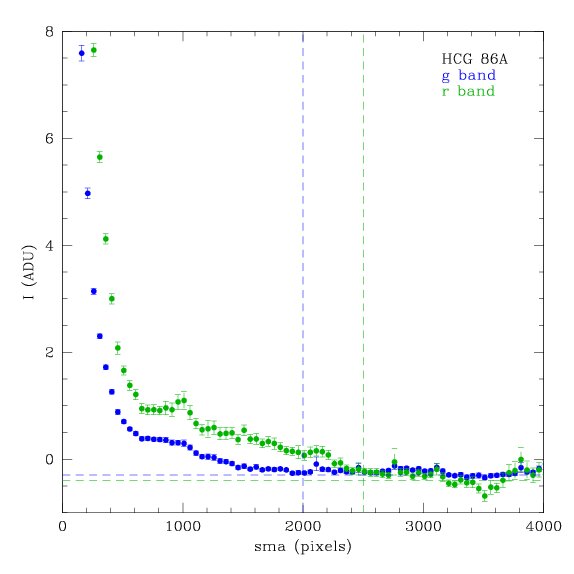}}
    \resizebox{0.5\textwidth}{.4\textheight}{\includegraphics{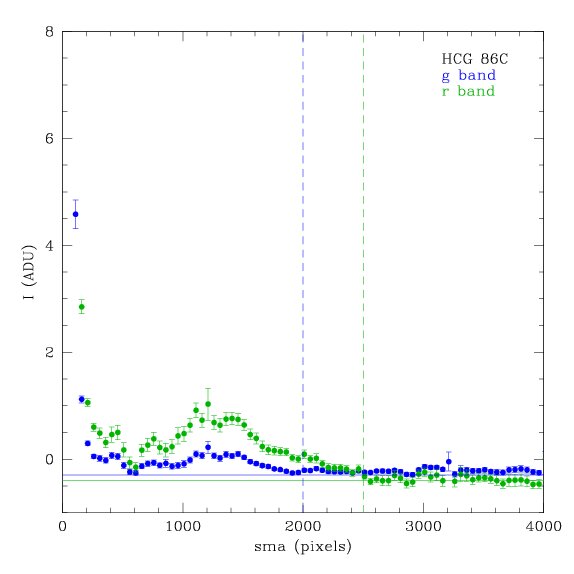}}
    \resizebox{.5\textwidth}{.4\textheight}{\includegraphics{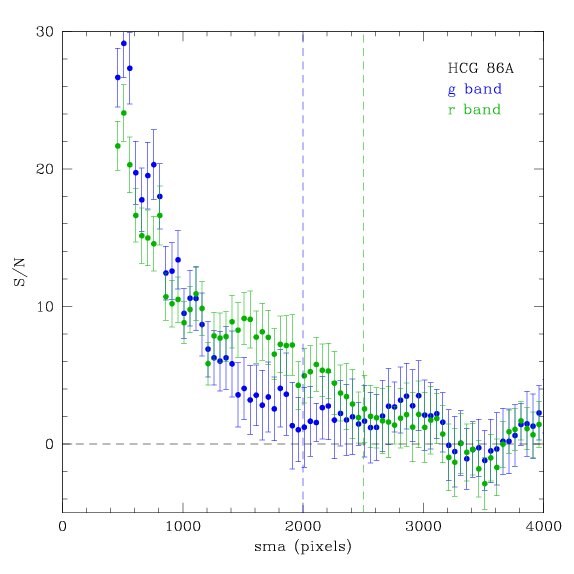}}
    \resizebox{.5\textwidth}{.4\textheight}{\includegraphics{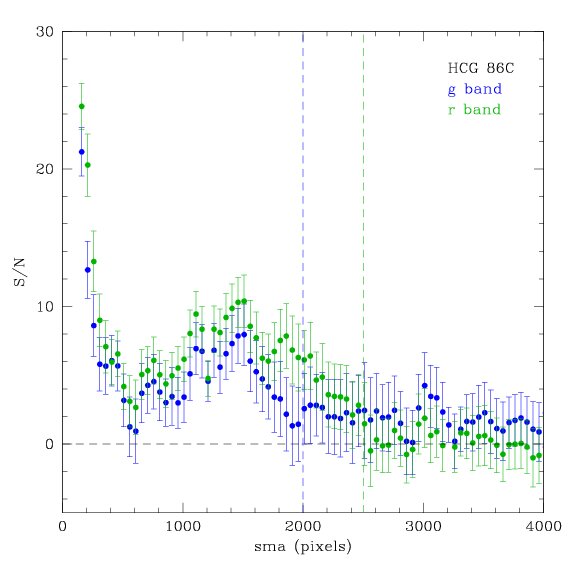}}
      \caption{{\it Top panels}: Intensity profiles for HCG~86A (left panel) and HCG~86C (right panel) in the $g$ and $r$ bands. The two vertical dashed lines show the estimated value for R$_{lim}$ for each band and the horizontal lines indicate the average value of the residual background fluctuations in each band.~ {\it Bottom panels}: Signal to noise ratio profiles for HCG~86A (left panel) and HCG~86C (right panel) in the $g$ (blue points) and $r$ (green points) bands. The vertical dashed lines indicate the R$_{lim}$ for each band and the horizontal lines indicate the zero value for signal to noise ratio. The bump observed at
      $\sim$ 1000-2000 pixels, in both $g$ and $r$ profiles centred on the HCG~86C, reasonably corresponds to the IGL contribution around the core of the group.} 
    \label{fig:limitrad}
\end{figure*}

In Fig.~\ref{fig:limitrad} we show the fitted isophote intensity as a function 
of the semi-major axis (sma) centered on HCG~86A (HCG~86B and HCG~86D share the same envelope as HCG~86A) and HCG~86C. 
Both fits provide consistent values for R$_{lim}$, 
with R$_{lim} =6.67$~arcmin in the $g$ band and R$_{lim} =8.33$~arcmin in the $r$ band.
Since the $i$-band images are shallower, the limiting radius is smaller, R$_{lim} =2$~arcmin, for both regions.
The S/N is shown in the lower panels of Fig.~\ref{fig:limitrad}.
At $R\geq R_{lim}$ from the galaxy center S/N$\sim1-3$ in both $g$ and $r$ bands, and 
the residual background fluctuations are $I_g=-0.29 \pm 0.05$~ADU, 
$I_r=-0.39 \pm 0.09$~ADU and $I_i=0.10 \pm 0.1$~ADU in the $g$, $r$ and $i$ bands, respectively. 
These values have been taken into account 
to compute the corresponding surface brightness magnitude limits and
error estimate\footnote{The total uncertainty on the surface brightness magnitudes takes into account the 
uncertainties on the photometric calibration ($\sim 0.003 - 0.006$~mag) and the RMS in the background fluctuations. 
They have been calculated with the following formula: 
$err= \sqrt{(2.5/(adu \times \ln(10)))^2 \times ((err_{adu}+err_{sky})^2) +  err_{zp}^2}$, where
$err_{adu}=\sqrt{adu/N-1}$, with N is the  number of pixels used
in the fit, $err_{sky}$ is the rms on the sky background and $err_{zp}$
is the error on the photometric calibration \citep{Capaccioli2015,Seigar2007}.}.

Once the R$_{lim}$ has been derived, we performed the fitting of the isophotes for all the 
galaxies of the group out to R$_{lim}$, where all the shape parameters (i.e. ellipticity and P.A.) are 
left free.
Since the ELLIPSE task does not allow us to account for the light distribution of all galaxies in a single run, 
we proceeded iteratively by fitting one galaxy at a time, adjusting the mask accordingly. 
This is described in detail in the next section.

\begin{figure*}

   \resizebox{.5\textwidth}{.4\textheight}{\includegraphics{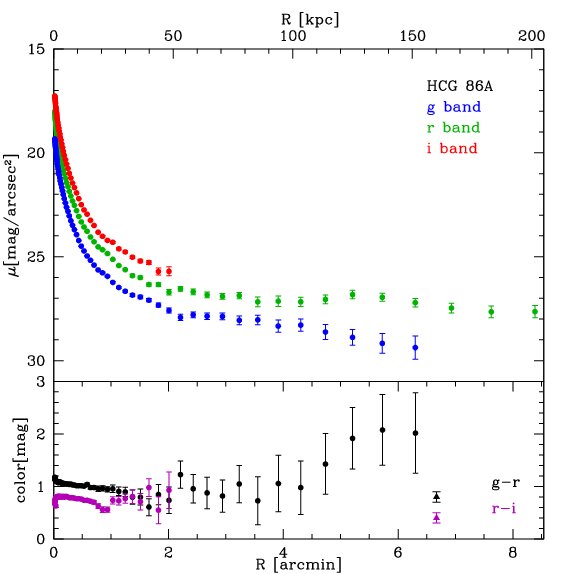}}
    \resizebox{0.5\textwidth}{.4\textheight}{\includegraphics{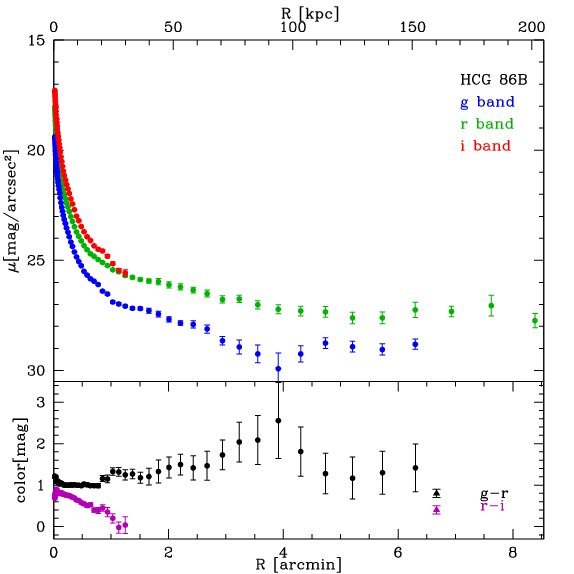}}
    \resizebox{.5\textwidth}{.4\textheight}{\includegraphics{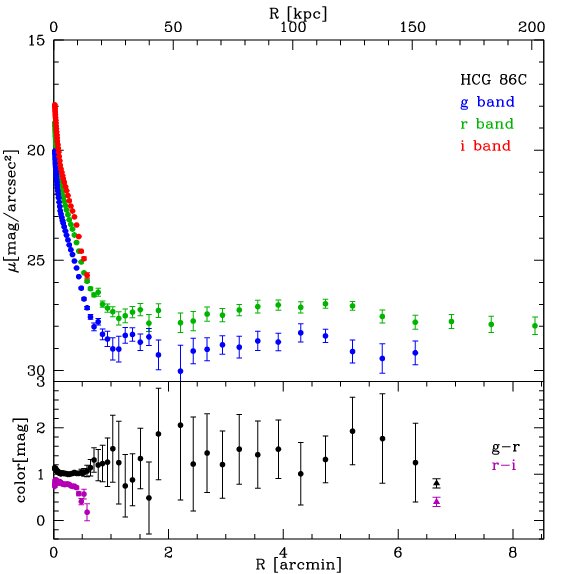}}
    \resizebox{.5\textwidth}{.4\textheight}{\includegraphics{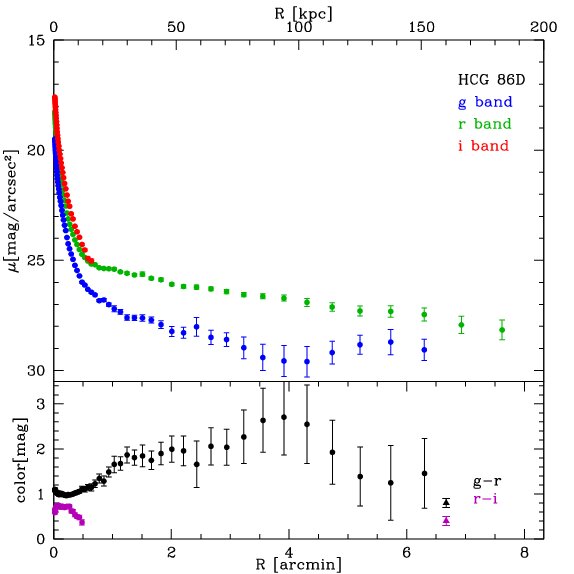}}
     \caption{{\it Top panels}: azimuthally-averaged surface brightness profiles of HCG~86A, HCG~86B, HCG~86C and HCG~86D in VST $g$ (blue), $r$ (green) and $i$ (red) bands. {\it Bottom panels}: $g-r$ (black) and $r-i$ (violet) color profiles of HCG~86A, HCG~86B, HCG~86C and HCG~86D. The points at 6.67 arcmin (i.e. $R_{lim}$ in the $g$ band) in each panel correspond to the integrated colors of the IGL component.} 
    \label{fig:prof}
\end{figure*}

\begin{table*}
\setlength{\tabcolsep}{1.2pt}
\centering
\caption{Structural parameters for the galaxies members of HCG~86 group from the fit of the isophotes.} 
\begin{tabular}{lcccccccccccc}
\hline\hline
     Galaxy     &    Morph. &     R.A &    Decl. &He-rad.&   $m_g$   & $m_g [R\leq R_{tr}]$ & $R_{e,g}$ & $R_{e,g}$  &  g-r  & r-i \\
     & Type & [J2000] & [J2000] &vel.[km/s]& [mag] &[mag] & [arcsec] &[kpc]& [mag] &[mag]\\
     (1)&(2)&(3)&(4)&(5)&(6)&(7)&(8)&(9)&(10)&(11)\\
    \hline
    $HCG~86A$ & E2 & 19:52:08.76& -30:49:32.70  & 6013 & 12.57 $\pm$ 0.06 &13.33$\pm$ 0.01& 130 & 52 &1.08 $\pm$ 0.02 & 0.79 $\pm$ 0.02\\
    $HCG~86B$ & E2 & 19:51:59.06& -30:48:58.40 & 5863 & 13.3 $\pm$ 0.2 &13.80$\pm$ 0.01& 287.7 & 115 & 0.98 $\pm$ 0.02& 0.77  $\pm$ 0.02\\
    $HCG~86C$ & SB0 & 19:51:57.46& -30:51:25.01 & 5317 & 14.60 $\pm$ 0.02&14.88$\pm$ 0.01& 7.67  & 3  & 1.12 $\pm$ 0.03& 0.84 $\pm$ 0.03\\
    $HCG~86D$ & S0 & 19:51:51.94& -30:48:30.4 & 5923 & 14.1 $\pm$ 0.1 &14.64$\pm$ 0.02& 25.62 & 10.25 & 1.00 $\pm$ 0.04 & 0.71 $\pm$ 0.03\\
    \hline
    \end{tabular}
    \tablefoot{Column 1 report the name of the HCG~86 group members. In Col 2 is given the morphological type. In Cols 3 and 4 are listed the celestial coordinates of each group member. In Col 5 is listed the heliocentric radial velocity. Velocities for HCG~86~A, HCG~86~B and HCG~86~D are from \citet{deCarvalho1997}, while that for HCG~86~C is from \citet{Jorgensen1995}. In Cols 6 and 7 are reported the total magnitude and the total magnitude at the transition radius  in g band, while in Cols 8 and 9 list the effective radius in the $g$ band, in arcsec and kpc respectively. The average $g-r$ and $r-i$ colors, derived up to transition radius, are listed in Cols 10 and 11. Magnitudes and colors were corrected for Galactic extinction using the extinction coefficients provided by \citet{Schlafly_2011}}.
    \label{tab:sample}
\end{table*}

\subsection{Fit of the galaxy light distribution}\label{subsec:fit}

In order to estimate the total amount of the IGL, the contribution to the 
light of the brightest inner regions of the galaxies in the group must be modelled and subtracted.
As mentioned in the previous section, this is an iterative process made of the following steps.
For each of the four brightest group members, starting with HCG~86A, and in each band, we have
{\it i)} performed the isophote fit (see Sec.~\ref{subsec:fitgal}) to derive the azimuthally-averaged surface brightness profile, out to R$_{lim}$;
{\it ii)} fitted the 1-dimensional (1D) profiles adopting  multi-component empirical laws to estimate the 
transition radius $R_{tr}$ between the brightest central region of the galaxy and the faintest outskirts; 
{\it iii)} derived the 2D model of the inner component at $R\leq R_{tr}$, using the IRAF task BMODEL, 
and subtracted it from the entire image. 
On this residual image we perform the isophote fit of each group member in turn.
A new mask is made at each step of the fitting analysis, where all contributions from background/foreground objects, 
cirrus emissions, residuals from subtracted stars are included. In addition, at each step, 
the residuals of the subtracted 2D model of the brightest regions of each galaxy is also masked and excluded from the fit.

The first fit was performed on the brightest group member, HCG~86A, since it dominates the light distribution.
The resulting azimuthally average surface brightness and color profiles for HCG~86A are shown in Fig.~\ref{fig:prof}, where R corresponds to the semi-major axis (sma) in arcmin.
We map the surface brightness profiles down to 
$\mu_g \sim 30$ mag/arcsec$^2$,  $\mu_r \sim 28$ mag/arcsec$^2$ and $\mu_i \sim 26$ mag/arcsec$^2$ in the $g$, 
$r$ and $i$ band, respectively.

To proceed with the isophote fit for all the other group members, it is necessary to model and subtract the 
light distribution of each galaxy in the group. To this aim, we need to derive the transition 
radius R$_{tr}$ between the brightest parts and the faint outskirts. This is derived by performing
the 1D decomposition of the galaxies' azimuthally-averaged surface brightness profiles. We adopted
the fitting procedure introduced by \citet{Spavone2017b}, also used in other VEGAS 
papers \citep[see e.g.][]{Spavone2018,Spavone_2020,Cattapan2019,Iodice2016, Iodice_2020}. 
This is motivated by several observational and theoretical works, which suggested that the stellar envelope 
in the galaxies' outskirts can be reproduced by adding an additional component to the S{\'e}rsic law, 
which fits well the inner and brightest regions of the galaxy
\citep{Seigar2007,Donzelli2011,Arnaboldi2012,Iodice2016,Spavone2017b,Spavone_2020}. 
As addressed in these studies, based on the photometry alone, the contribution of the diffuse light
cannot be separated from the stellar envelope \citep[e.g.][]{Gonzalez2007,Seigar2007}. 
However an estimate of the total contribution from the faint outskirts (i.e. stellar envelope 
plus diffuse light) can be disentangled from the bounded and bright emission in galaxies by fitting 
the total integrated light with analytic models.
This is particularly enhanced in compact groups (as HCG~86) where galaxies are so close in projection
that their stellar envelopes merge with each other, fully engulfing the intra-group diffuse light.

In all HCG~86 group members, we used a S{\'e}rsic law to model the main body of the galaxies 
and an exponential law to reproduce the diffuse component (stellar halo plus the IGL). 
The results of the fit are provided in Appendix~\ref{appA:1Dfit}. 
The best-fitting parameters are reported in Tab. \ref{tab:1Dfit}, including $R_{tr}$.

The 1D fit of HCG~86A shows an extended exponential component
in the outskirts, at $R\geq 51$~arcsec ($\sim 20.4$~kpc), with a central surface brightness of $\mu_0=26.30$~mag/arcsec$^2$ and a scale length $r_h=132$~arcsec ($\sim 53$~kpc) in the $g$ band.
Based on the 1D-fitting results, the 2D model for HCG~86A has been made for the brightest regions of the galaxy,
i.e. for $R \leq R_{tr}$=51~arcsec, and then subtracted from the image in each band.
Using the residual image, the other two galaxies in the core of the group, HCG~86B and HCG~86D, 
have been modelled in turn, i.e. the azimuthally-averaged 
surface brightness profiles are obtained from the isophotal analysis and then fitted. 
As expected, since they are all physically close in projection and completely embedded in a diffuse stellar
halo, the surface brightness profiles all show the extended outer exponential component (see Fig.~\ref{fig:prof}).
Based on the 1D fit, we have modelled and subtracted only the brightest regions of each galaxy (i.e out to their $R_{tr}$) and,
consistently, the remaining outer component has a scale length of the same order of magnitude and comparable
$\mu_0$ (see Tab.~\ref{tab:1Dfit}).
As a final step, on the resulting image where all the three galaxies in the core of the group 
have been modelled and 
subtracted, we have performed the analysis of HCG~86C, the S0 galaxy located to the SE of the group centre (see Fig.~\ref{fig:col_comp}). The surface brightness profiles for this object also show an extended envelope in $g$ and $r$ bands, 
as detected for the three galaxies in the core of the group (see Fig.~\ref{fig:prof}). 

The resulting final residual image where all the 2D models of the brightest regions of group members have 
been subtracted, is shown in the right panel of Fig.~\ref{fig:cirrus_IGLmap}.
This image is then used to estimate the total amount of IGL in the group, 
assumed to be the contribution of the stellar envelope plus the diffuse light in the intra-group medium.

\begin{table*}
\setlength{\tabcolsep}{1.2 pt}

\centering

\caption{Structural parameters derived from the 1D fit of the azimuthally averaged surface brightness profiles of the brightest group members.}
\label{tab:1Dfit}

\begin{tabular}{lcccccccccc}
\hline\hline
Object & $\mu_{e,g}$ &$R_{e,g}$ & $R_{e,g}$ & $n_{g}$ & $\mu_{0,g}$ & $r_{h,g}$ & $r_{h,g}$ & $R_{tr,g}$ &$R_{tr,g}$ \\ 
    & [mag/arcsec$^{2}$] &[arcsec]& [kpc] &&[mag/arcsec$^{2}$] & [arcsec]& [kpc]&[arcsec]& [kpc]\\
  (1)  & (2) &(3)& (4) & (5) & (6)& (7)& (8)& (9)& (10)\\
\hline \vspace{-7pt}\\
HCG~86A  & 22.2$\pm$0.1 & 11$\pm$2  & 4.4$\pm$0.8&1.92$\pm$0.06&26.30$\pm$0.01&132$\pm$1&52.8$\pm$0.4&51&20.4\\
HCG~86B  & 22.66$\pm$0.08 & 9.1$\pm$0.2  & 3.64$\pm$0.08&2.6$\pm$0.2&26.6$\pm$0.3&133$\pm$51&53$\pm$20&42&16.8\\
HCG~86C & 22.8$\pm$0.2 & 7.5$\pm$0.7 &3.0$\pm$0.3& 2.1$\pm$0.2&28.5$\pm$0.3&800$\pm$24&320$\pm$10&45&18\\
HCG~86D & 22.11$\pm$0.04 & 6.2$\pm$0.2  &2.48$\pm$0.08& 2.1$\pm$0.4&26.5$\pm$0.2&91$\pm$7&36$\pm$3&30&12\\
\hline
\end{tabular}
\tablefoot{Columns 2, 3, 4, 5 report the effective surface brightness, effective
radius (in arcsec and kpc scale) and S{\'e}rsic index for the inner component of each fit, in the g band, whereas columns 6, 7 and 8 list the
central surface brightness and scale length for the outer exponential component, in arcsec and kpc scale. Columns 9 and 10 give a transition radii, in arcsec and kpc scale respectively, derived by the intersection between the first and the second component of the fit \ref{appA:1Dfit}.}

\end{table*}

\section{Results}

For all the galaxies in the group, the isophote fits give the azimuthally-averaged surface brightness and color profiles as output, shown in Fig.~\ref{fig:prof}.
Because the $g$ and $r$ images are deeper than the $i$-band, the $g-r$ color profiles are more extended in radius 
(out to $\sim6$~arcmin) than the $r-i$ (out to $\sim 0.5-2$~arcmin). 
For the galaxies in HCG~86 
a gradient toward redder colors ($\sim 1.2 - 2$~mag) is observed in all the $g-r$ profiles at $R\geq 2-4$~arcmin.
As discussed in detail later in Sec. ~\ref{sec:cirrus}, the region where HCG~86 group resides
is contaminated by the Galactic cirrus emission. Therefore, we suspect that the observed reddening is 
due to this contamination, which is stronger in the $r$ band with respect to the $g$ band \citep[][]{Rom_n_2020}.
This change in slope in the color profile was observed by \citet{Watkins_2016} for the spiral galaxy M64, 
where the $g-r$ color profiles of the galaxies get redder with radius, and is addressed by cirrus contamination. 
This effect does not occur in the $r-i$ color since the cirrus in these bands has a comparable emission 
\citep[for details see also Fig.2 of][]{Rom_n_2020}, even though the $r-i$ color profile of the BGG (see top panel in 
Fig.~\ref{fig:prof}) becomes shallower at $R\sim 0.9$ arcmin. As described in detail in the following section, this 
corresponds to the transition radius of the HCG~86A, beyond which the IGL component begins to be present.  
A similar trend in the $r-i$ color profile was also found in several works \citep[e.g.][]{Zibetti2005, 
coccato2008,Greene2015}. In particular, for the Abell Cluster 85, \citet{montes2021buildup} suggested that the reddening in $r-i$ color profile, at larger radii,
could be an observational evidence of ICL as accreted stars on the BCG.

Finally, from the azimuthally-averaged surface brightness profile we have derived the total magnitude 
for each group member and, therefore, the
integrated $g-r$ and $r-i$ colors. These are reported in Tab.~\ref{tab:sample}. 
The total luminosity of the four brightest group members, which includes the brightest central parts, the stellar envelope and IGL, is $L_g^{TOT} = 1.76 \times 10^{11} L_{\odot}$ and $L_r^{TOT} = 2.8 \times 10^{11} L_{\odot}$, in the $g$ and $r$ bands respectively.

\subsection{The intra-group light in HCG~86}\label{sec:igl}

\begin{figure*}
    \centering
    \includegraphics[width=18cm]{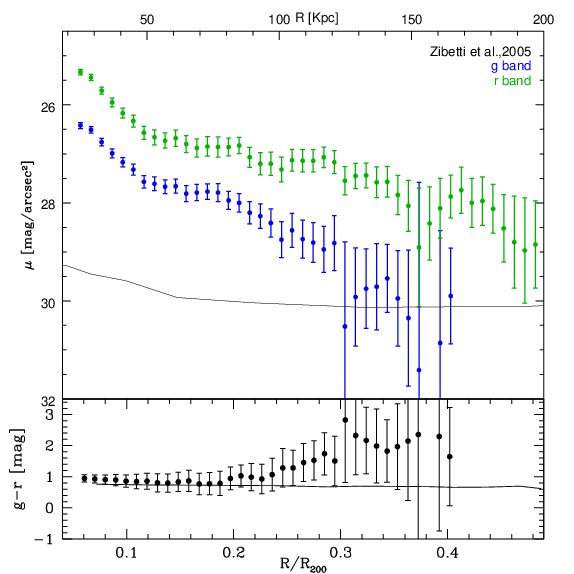}
     \caption{{\it Top panel}: Azimuthally averaged surface brightness profiles of the IGL in the HCG~86 group derived in $g$ and $r$ bands (blue and green points, respectively). The solid black line is the average surface brightness profile of the IGL for galaxy clusters by \citet{Zibetti2005}.
     {\it Bottom panel}: Azimuthally averaged $g-r$ color profile of the ICL in the HCG~86 group compared with the range of $g-r$ colors given by \citet{Zibetti2005} for ICL in galaxy clusters. The group centric and cluster centric distances are normalized for their R$_{200}$.}
     \label{fig:mu_igl}
       \end{figure*}

The residual image, obtained by subtracting the models of the brightest stars and galaxies of the group (see Sec.~\ref{subsec:masking} and Sec.~\ref{subsec:fitgal}), derived in the $g$ band, is shown in Fig.~\ref{fig:cirrus_IGLmap}. This image reveals the distribution of the diffuse intra-group light in HCG~86.
The HCG~86 group is fully embedded in an extended 
envelope of diffuse light, indistinguishable from the stellar halos of the brightest galaxies,
elongated in the West-East direction, like the distribution of the members of the group.
Several faint filamentary structures, also noted in the original image in Fig.~\ref{fig:HCG86_dss}, are present in the Southern regions of HCG~86A, where the 
contamination from the cirrus's light is at a minimum (see Sec.~\ref{sec:cirrus}).
From this image we have derived the azimuthally-averaged surface brightness profiles of the IGL, 
in both $g$ and $r$ bands, by fitting the light distribution in circular isophotes, fixing the centre in HCG~86A. 
These are shown in Fig.~\ref{fig:mu_igl}.
We adopted the same mask used for the surface photometry of the galaxies, where all bright sources 
(foreground stars and background galaxies) and the regions contaminated by cirrus emission 
(right North of the group and the filament to the West, see Fig.~\ref{fig:cirrus_IGLmap}) are excluded from the fit.
The $g$ and $r$ IGL surface brightness profiles extend out to the limiting radius estimated from the isophote fit for each band (see Sec.~\ref{sec:phot}), i.e. out to 6.67 arcmin ($\sim$ 160 kpc) and down to $\sim 30$~mag/arcsec$^2$ in $g$ band 
from the centre of HCG~86A, and out to 8.33 arcmin ($\sim$ 200 kpc) and down to $\sim 29$~mag/arcsec$^2$ in $r$ band 
from the centre of HCG~86A. We map the $g-r$ color profile of the IGL out to 160 kpc. 
This is the most extended estimate of the IGL derived in a group of galaxies. 
Previous observations find that the IGL in HCGs spans a range in radii from 30 to 80 kpc from the center of the group \citep{DaRocha2005,DaRocha2008}. 
Recent measurements by \citet[][]{Poliakov2021} map the IGL out to $R_e \sim 100$~kpc.

In the inner regions ($R\leq100$~kpc) the IGL profiles have an exponential decrease which
resembles the contribution of the stellar envelopes. At larger radii,
it is shallower and this might be the "pure" contribution of the diffuse and not-bound
light in the group. These components cannot be separated using photometry
alone, and deep spectroscopic observations are needed.
Therefore, as has been the custom for previous photometric studies of IGL and ICL using multi-component fits,
\citep{Seigar2007,Gonzalez2007,Zibetti2005},
we refer to the IGL as the contribution from the stellar envelope (see Sec. ~\ref{subsec:fit}) plus the diffuse light in the outskirts.
The IGL surface brightness profile for the HCG~86 group is compared with the average 
ICL profile derived by \citet[][]{Zibetti2005} for galaxy clusters (at z$\sim$0.25; see Fig.~\ref{fig:mu_igl}). 
In the region where the intra-cluster light dominates (at R$\geq 100$~kpc), both distributions have a 
comparable surface brightness ($\mu_g\sim 29 - 31$~mag/arcsec$^2$) in the $g$ band.
At smaller radii, the IGL surface brightness profile is mainly dominated by the light distribution of the stellar
envelopes around the group members ($\sim$ 26.5-28.5 mag/arcsec$^2$ in the $g$ band). 
In the bottom panel of Fig.~\ref{fig:mu_igl} we show the azimuthally- averaged $g-r$ color profile for the IGL in HCG~86. 
It is constant at $g-r\sim0.75$~mag for $R\leq120$~kpc. At larger radii, the colors tend to be redder,
as also observed in the color profiles derived for each galaxy of the group (see Fig.~\ref{fig:prof}).
As discussed in Sec.~\ref{sec:cirrus}, this could be due to contamination from cirrus.
An average $g-r\sim0.75$~mag for the IGL in HCG~86A is comparable with the average value derived by
\citet[][]{Zibetti2005} for $\sim$ 600 stacked galaxy clusters (scaled at the redshift 
of HCG~86, see Fig.~\ref{fig:mu_igl}). Moreover, such a value is consistent with the $g-r$ colors derived in other groups and galaxy clusters, where $g-r=0.68$~mag in the Abell Cluster 2744 at z=0.3 \citep{Montes2014}, $g-r\sim 0.7$~mag in the Fornax Cluster \citep{Iodice2017a, Raj_2020} and also consistent with $g-i\sim 1.3$~mag found in the Abell Cluster 85 \citep{montes2021buildup}.

Since we consider that the IGL profile could be contaminated by cirrus emission at $R\leq120$~kpc, we have derived two different estimates of the total integrated flux for the IGL in HCG~86, in the $g$ and $r$ bands.
In a circular area centred on the BGG (HCG~86A), a conservative value is derived within $R\leq120$~kpc, 
where the surface brightness and colors are comparable with previous studies, and a second value has been obtained by
including the whole IGL profile out to 160~kpc.
The extinction-corrected\footnote{The IGL magnitudes and color were corrected for Galactic extinction using the extinction coefficients provided by \citet{Schlafly_2011}.} 
magnitude of the IGL within 120 kpc is $m_g$= $ 13.9 \pm 0.1$~mag 
in the $g$ band, with an average color of $g-r = 0.83 \pm 0.3$ mag.
Within this radius, {\it i)} the total luminosity of the IGL is $L_g^{IGL}= 2.80 \times 10^{10} L_{\odot}$ and
$L_r^{IGL}= 3.02 \times 10^{10} L_{\odot}$ in $g$ and $r$ bands respectively, and {\it ii)} the fractions of IGL with 
respect to total luminosity of the group (given in Sec.~\ref{sec:phot}) 
are $\simeq 16 \pm 3\%$ and $\simeq 11 \pm 2\%$ in the $g$ and $r$ bands, respectively.
Since the total luminosity of the BGG HCG~86A is $L_g^{A} = 
9.55\times 10^{10} L_{\odot}$ in the $g$ band and $L_r^{A} = 1.29 \times 10^{11} L_{\odot}$
in the $r$ band, the IGL fraction with respect to the BGG is $\sim28 \pm 5\% $
and $\sim23 \pm 7\%$ in the $g$ and $r$ band, respectively.

At $R<160$~kpc, the extinction-corrected
magnitude of the IGL is $m_g$= $13.7 \pm 0.2$~mag in the $g$ band, with an average color of $g-r = 0.85 \pm 0.3$ mag.
The total luminosity of the IGL is $L_g^{IGL}= 3.35 \times 10^{10} L_{\odot}$ and
$L_r^{IGL}= 3.79 \times 10^{10} L_{\odot}$ in $g$ and $r$ bands respectively.
The fractions of IGL with respect to the total luminosity of the group (given in Sec.~\ref{sec:phot}) are $\simeq 19 \pm 3\%$ and $\simeq 14 \pm 2\%$ in the $g$ and $r$ bands, respectively. The two estimates of the IGL fraction provided above are consistent within the uncertainties and are within the estimated contamination of the cirrus (see Sec.~\ref{sec:cirrus}). The fractions of IGL, at $R<160$~kpc, with respect to the luminosity of the BGG are  
$\sim35 \pm 5\%$ and $\sim29 \pm 6\%$ in the $g$ and $r$ bands, respectively.  By photometric dissection with double S{\'e}rsic decomposition, \citet[][]{Kluge2021} found a ICL/BCG fraction in a range of 31 $\%$< ICL/BCG < 73 $\%$, which is consistent with our results.
 The BGG/(BGG+IGL) fraction in HCG~86 is equal to 60$\%$, in agreement with the simulations performed by \citet[][]{Contini21} for a halo with $M_{vir}\sim 10^{13}$~M$_{\odot}$.

The average colors derived for the IGL in HCG~86 are also comparable with values published by \citet[][]{DaRocha2005} and \citet[][]{DaRocha2008}, for their detailed analysis 
of the IGL in compact groups. 
They found that the B-R colors for the IGL range from  $0.85-1.75$ mag, which corresponds 
to $g-r \simeq 0.3-0.8$~mag, adopting the 
color transformation from \citet[][]{Kostov2018}. 

\subsection{Cirrus contamination}\label{sec:cirrus}

As reported in the Sec.~\ref{subsec:cirrus_mask}, the field around HCG~86 is affected 
by the contamination of Galactic cirrus. In order to provide an accurate estimate of the IGL, 
we need to quantify the amount of contamination from cirrus.
From the $100\mu m$ map we have estimated the minimum and maximum values of 
the flux along the filament to the West, which is assumed to be pure cirrus emission. 
These values range from  9.25 to 9.78~MJy/sr. 
According to the location of HGC~86 in the $100\mu m$ map, the group resides in a 
region where the cirrus emission is
about  $9.30 ~ MJy/sr$, so very close to the minimum (see Fig.~\ref{fig:cirrus_IGLmap}). 
Therefore, we expect that the contribution of the cirrus' light to the IGL is low.

\begin{figure*}
    \centering{}
    \includegraphics[width=16cm]{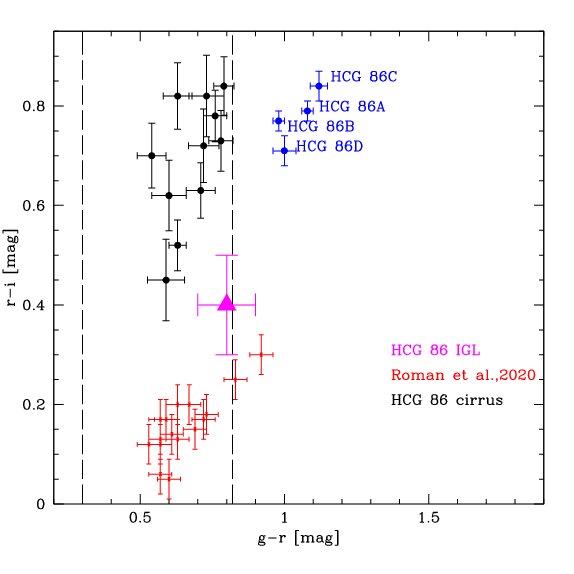}
    \caption{$g-r$ versus $r-i$ color-color diagram. Colors of the cirrus derived in the West filament are in black. The integrated colors of the group members in HCG~86 
     are in blue. The average value of the IGL in the group is  shown as magenta triangle (see Sec.~\ref{sec:igl}). The small red points are the colors derived for Cirrus from \citet[][]{Rom_n_2020}. The two vertical dashed lines indicate the range of $g-r$ colors of the IGL in the HCGs studied by \citet[][]{DaRocha2005,DaRocha2008}.In this color-color plot the region in which the IGL lies is different both from those typical of cirrus \citep[][]{Rom_n_2020} and from cirrus in the HCG~86 space. This confirms that IGL in HCG~86 are not largely contaminated by cirrus}.
    \label{fig:col_cirrus}
\end{figure*}

Unfortunately, when dealing with Galactic cirrus, a direct decontamination of optical images is not possible \citep{mihos2019deep}.
As suggested by \citet{Rom_n_2020}, a multi-wavelength approach can help to discriminate between Galactic cirrus emission and real LSB features. 
To achieve this, we have used the cirrus contour levels from the IR map to define the regions on the optical images 
where cirrus contributes. In these regions, we have derived the $g-r$ and $r-i$ integrated colors of 
the cirrus.
Since the $R_{lim}$ estimated in the $g$ and $r$ bands are $6.67$~arcmin and $8.33$~arcmin, respectively (see 
Sec.~\ref{subsec:fitgal}), it is reasonable to assume that the emission from the western filament, which is further 
from the center of the group ($\sim 8-10$ arcmin), comes from cirrus.
In this region, from the optical VST $g$, $r$ and $i$ images, we have derived 
the $g-r$ and $r-i$ colors in several circular areas along the filament and they 
are shown in Fig.~\ref{fig:col_cirrus}, by masking the contribution of all foreground stars and background objects.
In this color-color plot, we have also included 
the colors of the inner brightest parts of the group members, derived inside the 
transition radius (see Sec.~\ref{subsec:fitgal}), and the average colors of the IGL 
estimated at larger radii, i.e. $R\geq R_{tr}$, where this component starts to 
dominate with respect to the bright central regions of the galaxies.
In agreement with \citet{Rom_n_2020}, we found that the optical colors of the 
cirrus are distinct from those of extra-galactic sources, i.e. the group members: 
cirrus is on average bluer 
($g-r \sim 0.4-0.8$~mag) than the galaxies ($g-r \sim 1-1.15$~mag,  estimated up to their transition radius).
Even considering the larger uncertainties in the color estimate, due to the low 
signal-to-noise in the $i$-band images, 
the IGL is also distinct from the cirrus in the color-color plane, ($g-r \sim 0.8$~mag and $r-i \sim 0.4$~mag). 

Since in the color-color plot the IGL colors are different from those 
typical of cirrus, this confirms that the region where the group and its IGL 
envelope are located are not largely contaminated by cirrus, 
therefore, we are confident that the IGL estimate is robust.

Using the empirical relation by \citet[][]{Rom_n_2020}, 
where $g-r$ = ((0.56 $\pm$ 0.06) $\times$ log(flux$_{100\mu m}$)) + (0.21$\pm$ 0.05),
we quantify the amount of contamination from cirrus, to be taken into account in the 
IGL uncertainties.
From the VST $r$-band image, along the cirrus filament we have derived the average 
flux in the same circular regions used to estimate the colors of the cirrus (showed in Fig.~\ref{fig:col_cirrus}).
Assuming that in the region where the group resides the $g-r$ colors 
of the cirrus are on average comparable with those in the filament (reported above), 
using the relation from \citet{Rom_n_2020}, we derived the flux in the $g$ band 
corresponding to the 100-$\mu m$ flux in the region where the group centre is located (see left panel of Fig.~\ref{fig:cirrus_IGLmap}).
This quantity, which is about $6.16 \times\ 10^5 counts/px$, is considered the total 
amount of the flux from cirrus that can contaminates the IGL flux. 
Taking into account the flux of the cirrus emission in the region of the group, obtained with the procedure 
explained so far, the contamination of this source to the total flux of the IGL (see Sec.~\ref{sec:igl})
is about 10\%, in both the $g$ and $r$ bands.

\section{Discussion} \label{sec:concl}

In this section we focus on the comparison of the IGL fraction and colors in HCG~86 with 
{\it i)} other observation estimates for the
diffuse light in groups and clusters of galaxies, and {\it ii)}
theoretical predictions, in order to constrain the formation process for this component.

\subsection{How does the IGL amount in HCG~86 compare with other observational estimates for diffuse light?}

In Fig.~\ref{fig:vir_mass} we compare the IGL fraction as a function of the virial mass for HCG~86, with the estimates 
obtained from VST images, based on the same methods and analysis, for loose groups of galaxies \citep[from][]{Spavone2018,Iodice_2020}, for the Fornax cluster \citep[][]{Spavone_2020} and for the FornaxA subgroup \citep[][]{Raj_2020}. In addition, in this plot we have also included the available IGL estimates 
for other compact groups of galaxies \citep[from][]{DaRocha2005,DaRocha2008, Pildis_1995,Poliakov2021} and the ICL in Coma and Virgo clusters of galaxies
\citep[][]{Mihos2017,Jim_nez_Teja_2019}. Although these estimates are all obtained with the same approach, it must be taken into account that some of them come from different photometric bands.
%
On average, the IGL fraction found in compact groups of galaxies ranges between 15\%-46\%, 
therefore the 16-19\% value obtained for HCG~86 is fully consistent with the existing estimates.
Using different approaches and tracers, other estimates for the IGL amount available in the literature are provided by \citet{Aguerri_2006}, who find a fraction of IGL equal to ~4.7$\%$ for HCG~44, using Intra Group planetary nebulae (IGPNe).
Fig.~\ref{fig:vir_mass} also shows that groups and clusters of galaxies with similar virial mass
($\sim 10^{13}$~M$_{\odot}$) show different total amounts of diffuse light, ranging from low fractions of
 10\%-20\% (as observed in the triplet NGC~1533, HCG~86, HCG~35 and in the 
Virgo cluster) to high fractions $\sim 40\%$ (in NGC~5018 loose group of galaxies, HCG~90, HCG~79, HCG~94 and the Coma cluster).
In the literature data, HCG~79 is the compact group with the highest IGL fraction of 46\%.
Other literature estimates for the fractions of ICL range 
from 10\% to 40\% going from groups to clusters \citep[e.g.][]{Feldmeier2004,Zibetti2005,McGee2010,Toledo2011}.

\subsection{Diffuse light versus virial mass}

As already noted by \citet[][]{Iodice_2020}, the large scatter observed in Fig.~\ref{fig:vir_mass} might indicate that there is no significant relationship between the diffuse light content and the virial mass of the environment, since 
large ICL fractions (~30-45\%) are observed in groups with $M_{vir}\sim 10^{13}$~M$_{\odot}$ and in massive clusters of 
galaxies like Fornax (with $M_{vir}\sim 10^{14}$~M$_{\odot}$) and Coma (with $M_{vir}\sim 10^{15}$~M$_{\odot}$).
To date, there is no general agreement in the literature regarding the relationship between $M_{vir}$ and ICL fraction \citep[see review by][]{montes2019intracluster}. 
On the observational side, \citet{Sampaio-Santos2021} found that the surface brightness of the diffuse light, in a sample of 528 clusters at 0.2 < z < 0.35, shows an increasing dependence on cluster total mass at larger radius. On the other hands,  \citet{Zibetti2005} find a constant fraction of ICL as a function of the halo mass, consistent with our results from Fig.~\ref{fig:vir_mass}.
On the theoretical side, several works have also found that the relationship between ICL and virial mass is flat and the fraction of diffuse intra-cluster light ranges between 20\% and 40\% 
\citep{Sommer-Larsen2006,Monaco_2006,Henriques_2010,Rudick2011,Contini2014}. These results would suggest that 
the driving factor for the IGL formation would not be related to the virial mass of the group or cluster environment \citep{Canas2020}. 

In contrast, \citet{Purcell2007} found a slight increase in the ICL fraction with the mass of the halo, 
from 20$\%$ to $30\%$ from halos with $M_{vir} = 10^{13} $~M$_{\odot}$ to those of $M_{vir} = 10^{15} $~M$_{\odot}$.
Also \citet{Lin_2004} and \citet{Murante2007} find that the ICL fraction increases weakly with cluster mass in their simulations.
Given such a large scatter in the observations, any definitive conclusion cannot be addressed at this point.

\subsection{Diffuse light versus ETGs-to-LTGs ratio}
\begin{figure*}
   \includegraphics[scale=0.63]{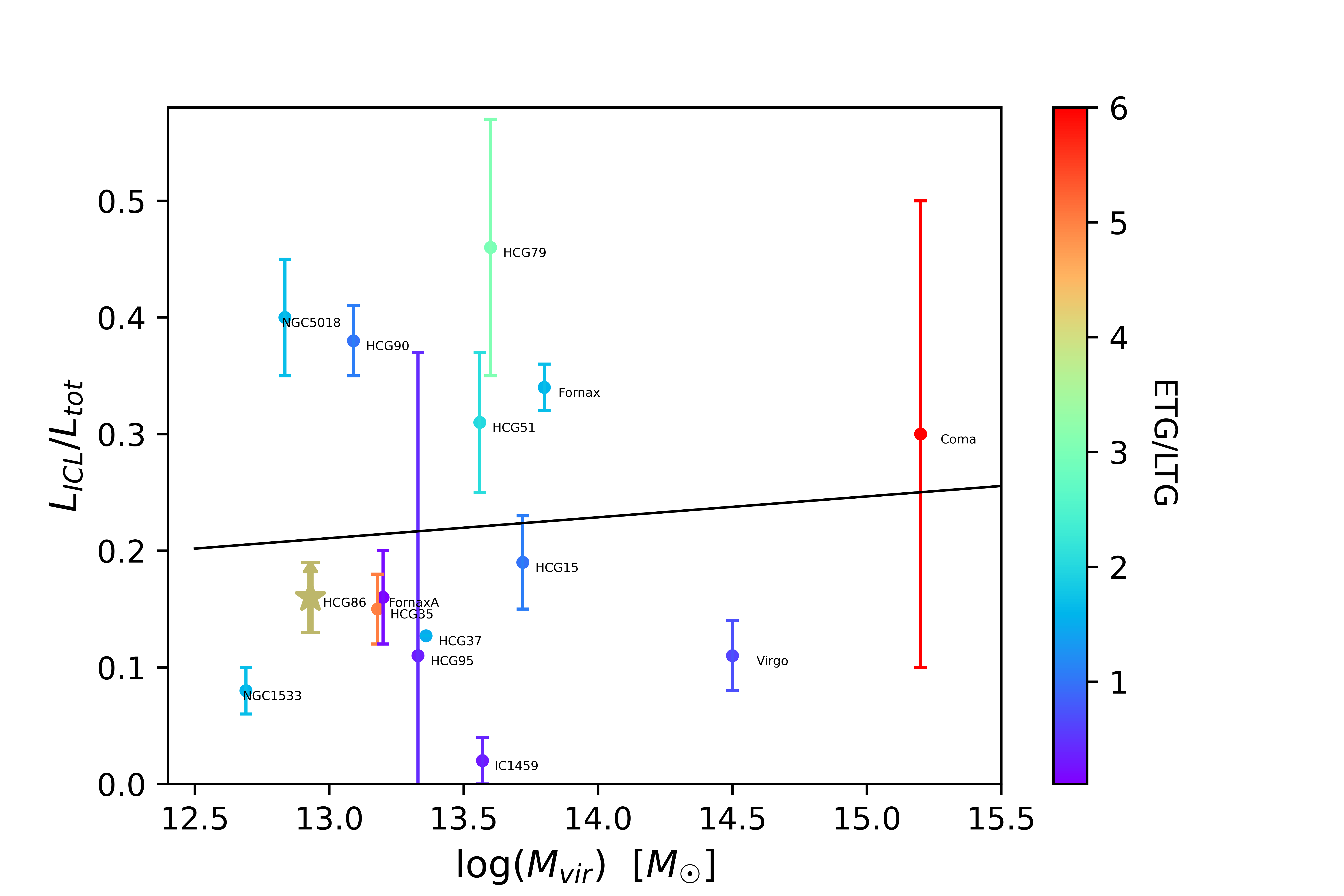}
   \includegraphics[scale=0.62]{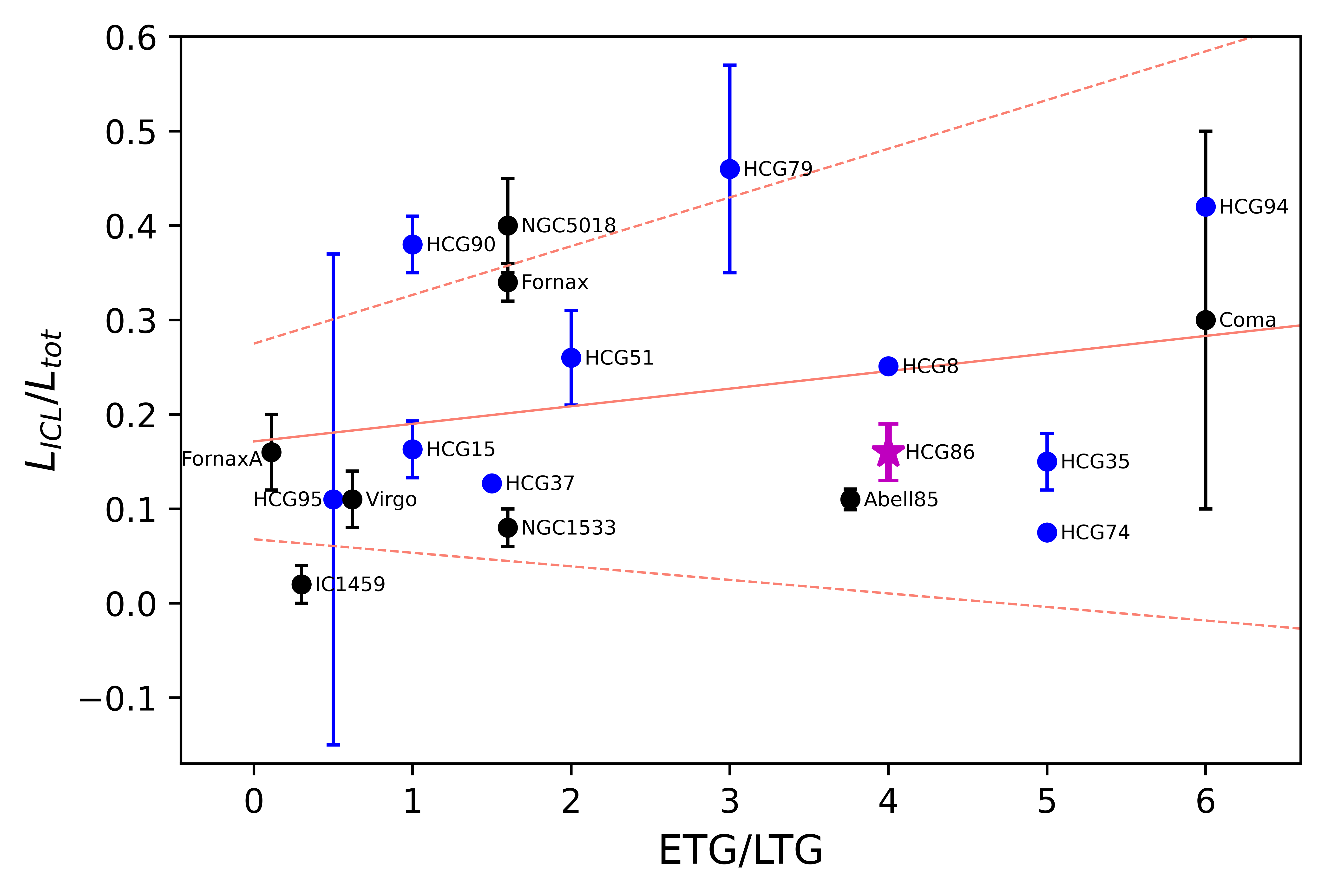}
    \caption{{\it Left panel:} Luminosity of the ICL/IGL component normalised to the total cluster/group luminosity as a function of the virial mass. 
    {The value derived for HCG~86 is compared with those for several Hickson Compact Groups, taken from \citet{DaRocha2008,Selim2008,Pildis_1995,Poliakov2021}. The estimate for HCG~94 and HCG~37 are provided by \citet{Pildis_1995} and \citet{Poliakov2021} without an error range, respectively.} In addition, values for the IGL estimated for other groups of galaxies derived using VEGAS data are also shown. These are: NGC~5018 group \citep{Spavone2018}, NGC~1533 triplet, IC~1459 group \citep{Iodice_2020}, and Fornax A subgroup \citep{Raj_2020}. The ICL fraction for the Fornax cluster is derived by \citet{Spavone_2020} using FDS data. We also report the values for Virgo \citep{Mihos2017}, Coma \citep{10.1093/mnras/180.2.207,Thuan_1977, Bernstein_1995, Adami_2005, Jim_nez_Teja_2019} and Abell~85 clusters \citep{Brough2017ApJ...844...59B, montes2021buildup}. Although these estimates are all obtained with the same approach, it must be taken into account that some of them come from different photometric bands.} The color of each point is coded according to its ETGs-to-LTGs ratio. The black line corresponds to the best fit for the linear correlation and it seems to suggest that any evident trend exists between the amount of ICL and M$_{vir}$. {\it Right panel:} Luminosity of the ICL component normalised to the total cluster or group luminosity as a function of the ETGs-to-LTGs ratio. 
    The IGL for all HCGs available in literature are marked as blue points. Other estimates for groups and 
    clusters of galaxies are indicated with the black points. 
    The value found in this work for HCG~86 is marked with magenta star-like point. 
    The coral solid line corresponds to the best fit for the linear correlation and the dashed coral lines mark 
    the 2$\sigma$ significance range of the correlation. The values for HCG~74, HCG~8, HCG~37, and HCG~17 (the latter overlaps to HCG~86) are from \citet[][]{Poliakov2021}. 
    \label{fig:vir_mass}
\end{figure*}

\citet{DaRocha2008} suggested that a high fraction of IGL is expected for groups dominated by early-type galaxies (ETGs), like compact groups.
In the left panel of the Fig. \ref{fig:vir_mass} we have color-coded the IGL fraction in groups and clusters based on their ETG to late-type galaxy (LTG) ratio. In right panel we show the fraction of the diffuse light with respect 
to the total luminosity of the cluster or group as a function of the ETGs-to-LTGs ratio.
In this figure we have included recent values published by \citet[][]{Poliakov2021} for five HGCs. 
Compared to HCGs of similar IGL fraction (HCG~35, HCG~95, HCG~15 and HCG~37), the ETG-to-LTG ratio in HCG~86 is quite
large (see right panel of Fig. \ref{fig:vir_mass}). This is four, since it is made up of only elliptical and S0 galaxies. 
HCG~17 has an IGL fraction ($\sim 16\%$) and ETG-to-LTG ratio similar to those deroved for HCG~86.
The NGC~5018 group has an ETG-to-LTG ratio similar to the NGC~1533 group, but the amount of IGL in the former structure is double with respect to the latter. The same is observed for IC~1459, Fornax~A, HCG~95 and HCG~15, all with comparable ETG-to-LTG ratios but different IGL amount ($\sim$2\% in IC~1459 and $\sim$16\% in Fornax~A).
Even though the large scatter, a weak trend between the amount of intra-cluster light and the ETG-to-LTG ratio seems to be present. More estimates for the IGL are needed to draw any definitive conclusion about this correlation.

\subsection{What could be the origin of IGL in HCG~86?}
As suggested by simulations \cite[e.g.][]{Contini2014,Contini2019}, the gravitational interactions between 
galaxies play an important role in the formation mechanism of the diffuse light in groups and clusters of galaxies.
As stated in Sec.~\ref{sec:intro}, these are predicted to be more efficient in groups of galaxies 
where the velocity dispersion of group members is low ($\sim$ 350 km s$^{- 1}$). 
In these simulations, from 5 to 25\%  of  the  diffuse  light builds from the infalling galaxies 
in the potential well of the BCG/BGG during the mass assembly history.
Most simulations predict that the bulk of the ICL is produced by the most massive 
satellite galaxies, M$_*$ $\sim$ 10$^{10 - 11}$~M$_\odot$ at lower redshifts, whereas
the contribution to the build-up of the diffuse light from the stripping of 
lower-mass galaxies (M$_*$ $\leq$ 10$^9$ M$_\odot$) is more efficient at higher ($z\sim1$) redshifts
\citep{Purcell2007,Contini2014,Martel2012, Contini2019}.

By comparing the IGL fraction and colors with simulations
from \citet[][]{Contini2014,Contini2019}, we address a possible origin for the IGL in HCG~86. It is worth noting that HCG~86  has $\log M_{200}\sim13$, which is close to the lower limit of their simulations.
Since \citet[][]{Contini2014,Contini2019} predict the fractions of mass in the ICL, we have derived the mass-to-light ratios (M/L)
corresponding to the $g-r$ and $r-i$ integrated colors derived for the IGL component and 
for the BGG (see Tab.~\ref{tab:sample}). 
In order to reproduce the ratio $L_{IGL}/L_{TOT}$ we gave for the light component, 
the integrated $g-r$ and $r-i$ colors for the BGG are obtained by considering the bright regions plus the faint outskirts.
Taking into account the error estimates on the colors, using the stellar population synthesis models based on the EMILES library \citep{Vazdekis2010},  we have derived the M/L in the following ranges:
$1.06\leq g-r_{BGG} \leq 1.10$~mag, $0.7 \leq g-r_{IGL} \leq 0.9$~mag, $0.77 \leq r-i_{BGG} \leq 0.81$~mag 
and $0.3\leq r-i_{IGL} \leq 0.5$~mag. 
Therefore, we have obtained that $M/L_{IGL}\sim 2-8$ and $M/L_{BGG}\sim 5-7$, with average values of 
$M/L_{IGL}\sim 5$ and $M/L_{BGG}\sim 6$.
Since the integrated colors of the other group members are comparable with that for the BGG (see Tab.~\ref{tab:sample} 
and Fig.~\ref{fig:col_cirrus}), it is reasonable to assume the same M/L ratio for all of them. 
As such, the $M_{IGL}/M_{TOT}$ $\sim14\%$, which is consistent with the 
estimates we based on the luminosity ratio of (16\% $\pm$ 3\%).

According to \citet[][]{Contini2014}, a fraction of IGL of about 20\%, comparable to 
that estimated in HCG~86, is consistent with a formation redshift within
$z\sim0.4-0.6$, which corresponds to a look back time of $\sim7-8$~Gyr
for formation of the group halo. These models predict that,
at this epoch and with this IGL fraction, disruption of satellite galaxies is the main 
channel to form the diffuse light. Tidal forces would give a higher fraction of IGL.
This scenario is consistent with the absence of any prominent bright tidal features 
or disturbed morphology in the core of HCG~86 and observing the IGL mainly in a diffuse form (see Fig.~\ref{fig:HCG86_dss}). The faint and diffuse bridge connecting HCG~86C to the rest of the group 
is the only evident tidal feature in the group, which would reconcile with the {\it late assembly} scenario proposed by
\citet[][]{Diaz-gimenez2021}, where HCG~86C might be the last member joining the group.

However, accretion events might have also contributed to the IGL
in HCG~86, but remnants (as tidal tails) of this process could have already dissolved, since their lifetime
is about 1 Gyr \citep[][]{Rudick2009,Mancillas2019}.
As a comparison, the HCG~90 and NGC~5018 groups, which are highly
interacting systems showing disturbed morphologies in the galaxies' outskirts and
the presence of tidal stellar tails, have larger fractions of IGL ($\sim40\%$).

From EMILES library, 
the IGL colors $g-r\sim0.7-0.9$~mag and $r-i\sim0.3-0.5$~mag are consistent with an age ranging from
7.8 to 11 Gyr (assuming a value for the metallicity (log(Z/Z$_{sun}$)) in a range of $[-0.4 , 0]$  , 
as predicted by \citet[][]{Contini2019} for the IGL at z=0). This interval would be even smaller towards lower 
ages considering that
the $r-i$ IGL color is an upper limit (see Fig.~\ref{fig:col_cirrus} and Sec.~\ref{sec:phot}).
Such an estimate is consistent with those based on the theoretical predictions 
discussed above.

Recent works have proved that ICL colors are a useful parameter to constrain the main 
formation process that contributed to the build up of this component \citep{Contini2019,Morishita2017,Montes2018,montes2021buildup}.
The range of values for the IGL colors of $0.7 \leq g-r\leq0.9$~mag in HCG~86 is also consistent with the range of  
$g-r$ colors predicted for the ICL by \citet[][]{Contini2019}, where $0.7\leq g-r \leq 0.8$~mag at z=0.
According to \citet[][]{Contini2019}, in this redshift range, the colors of the intra-cluster diffuse light are comparable with those of intermediate/massive galaxies 
($10^{10} \leq M_{*} \leq 10^{11}$~M$_\odot$), which therefore are the main contribution
to the mass of the ICL and IGL.
Since HCG~86 is dominated by ETGs, this would suggest that this group is quite evolved in
the mass assembly framework. Therefore, the existence of massive satellites that are merging
into the gravitational potential of the group members would be consistent
with the evolutionary phase of the whole system.

\section{Conclusions}\label{conc}

In this paper we presented deep images for the compact group of galaxies HCG~86 as part of the VEGAS survey.
The long integration time and wide area make our data deeper than previous literature studies of the IGL in compact groups of galaxies and allow us to detect IGL out to ($\sim 160$~kpc) and down to a surface brightness level of $\sim 30$~mag/arcsec$^2$
in the $g$ band. 

The main results are:

\begin{itemize}

    \item The IGL in HCG~86 is mainly in diffuse form, since we do not detect any bright 
    or extended tidal tails or stellar streams in the intra-group space. 
    The average IGL colors are in the range $0.7 \leq (g-r)_{IGL} \leq 0.9$~mag and $0.3\leq (r-i)_{IGL} \leq 0.5$~mag. 
    The $r-i$ color must be considered as an upper limit since the $i$-band image is shallower ($\sim 
    26$~mag/arcsec$^2$) than the other bands, therefore we are not able to map the 
    entire region of the IGL as in the $g$ and $r$ bands.
    
    \item The fraction of IGL in HCG~86 is about $\sim16$\% of the total luminosity of the group in the $g$ band, 
    and this is consistent with the same estimates available for other compact groups and loose groups of galaxies 
    of similar virial masses (see left panel in  Fig.~\ref{fig:vir_mass}).

    \item By comparing the amount of ICL with the cluster/group $M_{vir}$ (see left panel in Fig.~\ref{fig:vir_mass}) no strong correlation seems to be present. 
    On the other hand, according with previous studies, the ICL fraction seems to be weakly related to ETGs-to-LTGs ratio, where larger amount of intra-cluster light is found in more evolved structures, dominated by ETGs (see right panel in Fig.~\ref{fig:vir_mass}).
    
    \item We have estimated that the IGL fractions suffers from 10$\%$ of cirrus contamination. As shown in Fig.~\ref{fig:col_cirrus} the IGL is well separated from the cirrus area of the color-color diagram, suggesting that the estimate of the IGL colors is robust. 
    
    \item According to the theoretical models of \citet[][]{Contini2014}, the amount of 
    IGL in HCG~86 would be the result of the accretion from intermediate/massive satellites at an epoch of $z\sim0.4-0.6$ ($\sim 7-8$~Gyr). Such an age estimate is consistent with that derived to account for the average IGL $g-r\sim0.7$~mag and $r-i\sim0.4$~mag colors, using stellar population synthesis models.
    
    \item The $g-r$ IGL color is consistent with the range of $g-r$ colors predicted for the ICL by \citet[][]{Contini2019}, where  $0.7\leq g-r \leq 0.8$~mag at z=0. In this redshift range, the colors are consistent with the scenario where the main contribution to the mass of the ICL comes from the intermediate/massive galaxies ($10^{10} \leq M_{*} \leq 10^{11}$~M$_\odot$).\\
    
\end{itemize}

This work is a pilot project within the VEGAS survey aimed at studying the low-density 
and less massive environments of galaxy groups,
which are still relatively unexplored in the low-surface brightness regime.
In particular, with the upcoming VEGAS data, we plan to fill the gap at low virial 
masses in the IGL-M$_{vir}$ plane (see Fig.~\ref{fig:vir_mass}) where, to date, 
we still lack both deep observations and theoretical predictions. 
New observations and analysis will provide a comprehensive database of observables that can be directly compared with simulations on hierarchical mass assembly.

\begin{acknowledgements}

We thank the anonymous referee for their helpful comments on
the paper.
RR, MS and EI acknowledge financial support from the VST
project (P.I. P. Schipani).
EI acknowledges financial support from the European Union
Horizon 2020 research and innovation programme under the Marie Skodowska-Curie grant agreement n. 721463 to the SUNDIAL ITN network.
GD acknowledges support from CONICYT project Basal AFB-170002.
\end{acknowledgements}

\bibliography{hcg86}
\bibliographystyle{aa}
\begin{appendix} 
\section{Results of multi-component fits}\label{appA:1Dfit}

\begin{figure*}
   \resizebox{.5\textwidth}{.4\textheight}{\includegraphics{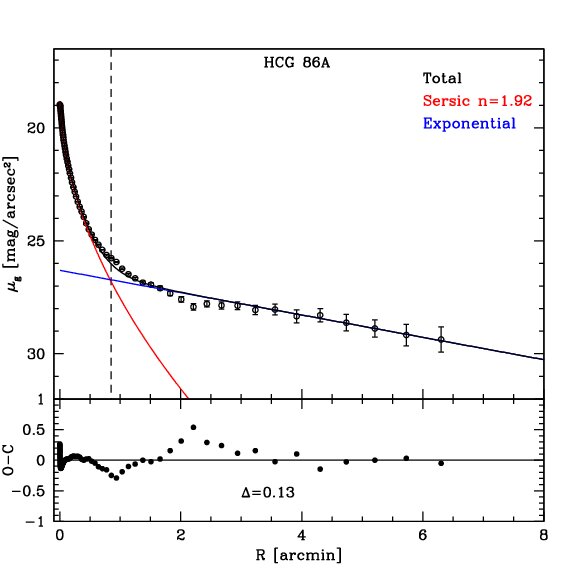}}
    \resizebox{0.5\textwidth}{.4\textheight}{\includegraphics{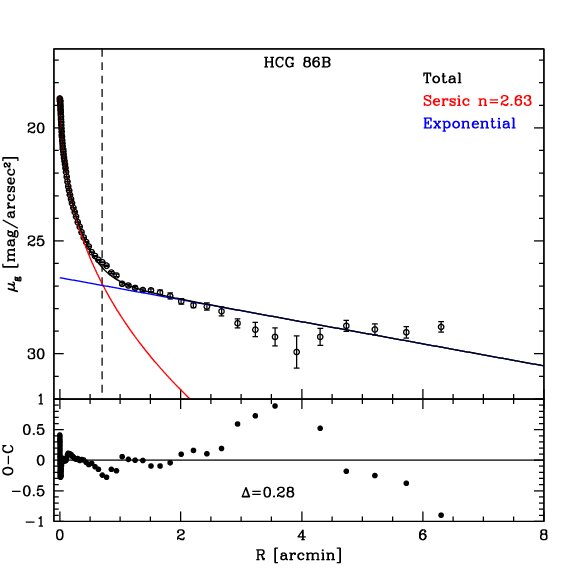}}
    \resizebox{.5\textwidth}{.4\textheight}{\includegraphics{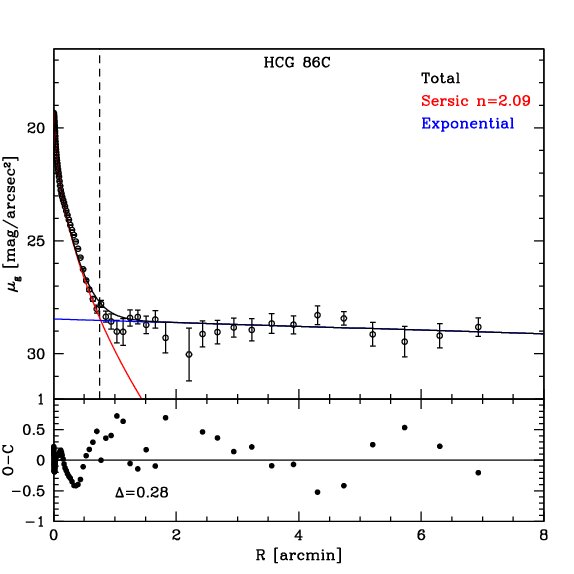}}
    \resizebox{.5\textwidth}{.4\textheight}{\includegraphics{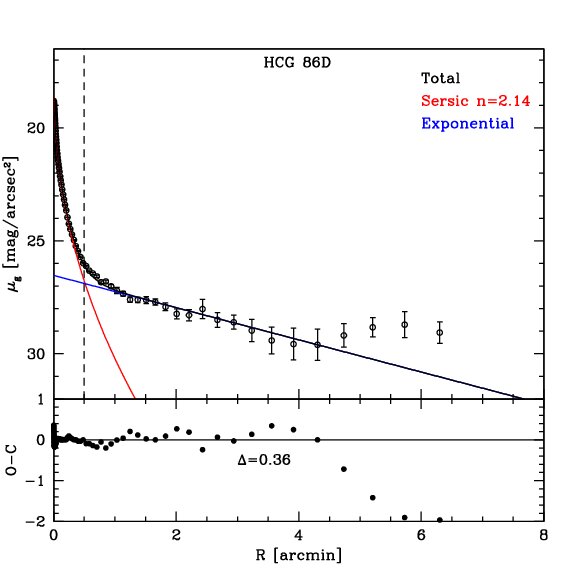}}
    \caption{Two components models of the azimuthally averaged surface brightness profiles of galaxies in HCG~86 group. The blue line indicates a fit to the outer diffuse component (halo+IGL). The red line indicates a fit to the inner regions with a Sérsic profile, and the black line indicates the sum of the components in each fit. The vertical dashed lines show the estimated value for $R_{tr}$ for each galaxy. {\it Bottom panel}: $\Delta$ rms scatter of the data minus the model (see text for details). }
    \label{fig:fit1d}
\end{figure*}

In this section we show the 1D multi-component fit performed on the azimuthally averaged
surface brightness profiles in the $g$ band, for all the four group members.
See in Fig.~\ref{fig:fit1d}.

\end{appendix}

\end{document}